\documentclass[a4paper,11pt]{article}
\pdfoutput=0 

\usepackage{jheppub} 

\usepackage[T1]{fontenc} 

\usepackage{epsfig,multicol
}
\usepackage{bm}
\usepackage{amsmath}
\usepackage{amsfonts}

\usepackage{wick}

\renewcommand{\Im}{{\rm Im \,}}

\def\slashchar#1{\setbox0=\hbox{$#1$}
   \dimen0=\wd0 \setbox1=\hbox{/} \dimen1=\wd1
   \ifdim\dimen0>\dimen1 \rlap{\hbox to \dimen0{\hfil/\hfil}} #1
   \else  \rlap{\hbox to \dimen1{\hfil$#1$\hfil}} / \fi}

\def\tr{\text{\rm tr}}
\def\Tr{\text{\rm Tr}}
\def\Det{\text{\rm Det}}
\def\det{\text{\rm det}}

\newcommand{\SU}{{\mathrm{SU}}}
\newcommand{\U}{{\mathrm{U}}}
\newcommand{\Eq}[1]{Eq.~(\ref{eq:#1})}
\newcommand{\eq}[1]{eq.~(\ref{eq:#1})}
\renewcommand{\neq}[1]{(\ref{eq:#1})}

\newcommand{\D}{\mathbf{D}}
\newcommand{\sD}{\slashchar{D}}
\newcommand{\sW}{\slashchar{W}}
\renewcommand{\L}{\mathcal{L}}
\newcommand{\R}{\mathbb{R}}
\newcommand{\C}{\mathbb{C}}

\newcommand{\mlr}{m_{LR}}
\newcommand{\mrl}{m_{RL}}
\newcommand{\thru}[1]{\slashchar{#1}}

\newcommand{\IaA}{a_1}
\newcommand{\IaB}{a_2}
\newcommand{\IbA}{b_1}
\newcommand{\IbB}{b_2}
\newcommand{\IbC}{b_3}
\newcommand{\IbD}{b_4}
\newcommand{\IbE}{b_5}

\newcommand{\gH}{g^-}
\newcommand{\gA}{g^+}
\newcommand{\OpH}{A^-}
\newcommand{\OpA}{A^+}
\newcommand{\EQ}{{}}
\newcommand{\EQb}{{}}
\renewcommand{\hm}{\hat{m}}
\newcommand{\hn}{\hat{N}}

\newcommand{\ignore}[1]{}

\title{\textsf{Leptonic CP violating effective action for Dirac and Majorana
    neutrinos}}

\author[]{Carmen Garc{\'\i}a-Recio and Lorenzo Luis Salcedo}

\affiliation[]{
Departamento de F\'{\i}sica At\'omica, Molecular y Nuclear and \\
  Instituto Carlos I de F\'{\i}sica Te\'orica y Computacional, \\
  Universidad de Granada, E-18071 Granada, Spain.}

\emailAdd{g\_recio@ugr.es}
\emailAdd{salcedo@ugr.es}

\abstract{\textsf{ 
In the Standard Model minimally extended to include massive neutrinos, we
compute the leading CP-violating zero temperature contributions to the
one-loop effective action induced by integration of the leptons. Such
contributions start at operators of dimension six and they are P even for
Dirac neutrinos and P even or odd for Majorana neutrinos. Dimension four
operators are allowed in the mixed Dirac-Majorana case. It is verified by
explicit calculation that CP can be violated in two generation settings for
Majorana neutrinos. Using different neutrino scenarios we give upper bounds
for the couplings of the CP-violating operators. As a rule, we find that
lepton-induced couplings are suppressed as compared to quark-induced
couplings, whenever the latter are allowed, nevertheless, through virtual
lepton-number violating mechanisms, Majorana neutrinos induce new CP-violating
operators not present in the quark or Dirac-neutrino cases.
}}


\date{today}

\begin{document} 
\maketitle
\flushbottom
\setlength{\unitlength}{1mm}


\sf

\section{\textsf{Introduction}}

Neutrino physics is nowadays quite an active field of research, from several
directions. These include nuclear physics (neutrinoless double beta decay,
matter effects, response functions of weak currents in nuclei), particle
physics (neutrino detection, neutrino oscillations, Standard Model extensions,
Pontecorvo-Maki-Nakagawa-Sakata (PMNS) matrix, CP violation, lepton number violation, sterile neutrinos),
astrophysics (neutrino production in stars, supernovae dynamics, neutrino
telescopes), cosmological (dark energy, inflation, primordial neutrinos), and
even more speculative ones such as using neutrinos for communication through
quantum channels. More importantly, these are not separated fields, rather
they are closely interconnected in such a way that advances in one fields
sheds light on all other fields as well \cite{%
Mohapatra:2005wg,
Bernabeu:1983yb,
Elliott:2002xe,
Morfin:2012kn,
GonzalezGarcia:2007ib,
Bertone:2004pz,
Halzen:2002pg,
Schwetz:2008er,
Hirata:1987hu,Bionta:1987qt,
Hinshaw:2012aka,
Stancil:2012yc
}.

On the other hand, CP violation remains as a challenging subject
\cite{Xing:2003ez,Buras:1997fb,Neubert:1996qg,Grossman:1997pa,Winstein:1992sx,%
  Paschos:1989ur,Wolfenstein:1987pe,Donoghue:1987wu,Charles:2004jd} since its
discovery fifty years ago \cite{Christenson:1964fg} and subsequent observation
of direct CP violation \cite{Burkhardt:1988yh,Aubert:2001nu}. CP violation
plays a key role in the understanding of baryo- and leptogenesis
\cite{Sakharov:1967dj,Trodden:1998ym,Ibrahim:2007fb}, time-reversal violation
(through CPT invariance) or the electric dipole moments of particles
\cite{Pospelov:2005pr}. There is no generally accepted explanation for the
non-violation of CP symmetry in the strong interaction sector
\cite{Cheng:1987gp}. In the electroweak sector, CP violation enters through
the flavor mixing complex mass matrices of the fermions, the
Cabibbo-Kobayashi-Maskawa (CKM) matrix for quarks \cite{Kobayashi:1973fv} and
the PMNS matrix for leptons
\cite{Pontecorvo:1957cp,Pontecorvo:1967fh,Maki:1962mu}. The CKM matrix
elements are currently known with some precision and the CP-violating phase
turns out to be rather small \cite{Beringer:1900zz}. For the PMNS matrix, the
angles are being measured in current experiments
\cite{Ahn:2012nd,Abe:2012tg,Abe:2013xua,An:2013uza} while no information is
currently available for the phases on which CP violation depends.

In the present work, we deal with the effective action of the Standard Model
(extended to include neutrino masses) and more concretely with its CP
violating component.  By effective action here we refer to the functional
obtained by integration of the fermions (quarks and leptons) in the
theory. Such functional depends on the configurations of the unintegrated
fields in the Standard Model, namely, the gauge bosons ($W^\pm$, $Z^0$, photon
and gluons) and the Higgs field. The effective action so defined is a
complicated gauge invariant functional of these bosonic fields. In order to
organize this functional we adopt a local expansion, namely, classifying the
terms by their number of covariant derivatives,
\begin{equation}
\Gamma = \int d^4x\, \sum_i g_i \, \mathcal{O}_i(x)
.
\end{equation} 
The quantities $\mathcal{O}_i(x)$ stand for the possible local operators
(monomials) that can be constructed using the available fields, restricted by
gauge and Lorentz covariance, etc. The $g_i$ are the couplings of these
operators in the effective action of the (extended) Standard Model.  Each
operator has a certain number of covariant derivatives. In this counting the
gauge fields count as one derivative, therefore (barring the Higgs field) the
operators are essentially classified by their mass dimension.\footnote{Of
  course, the mass dimension carried by the Higgs field is relevant for the
  (non) renormalizability of the operators. The Higgs field is properly
  included in our calculation below, we merely disregard it in our
  classification of operators.}  We aim at the computation of the couplings to
the leading (lowest dimensional) operators which are CP odd.  The effective
action has been modeled before in the literature, assuming phenomenological
values or estimates for the couplings to non-renormalizable operators, with
the purpose of studying electroweak baryogenesis or electric dipole moments of
particles
\cite{Shaposhnikov:1987pf,Dine:1990fj,Zhang:1993vh,Lue:1996pr,GarciaBellido:1999sv,Tranberg:2003gi}.
At variance with this phenomenological approach, our purpose here is to carry
out a direct calculation of the couplings using a strict derivative expansion
starting from the Standard Model Lagrangian.

The specific motivation for this calculation comes from the observation by
Smit \cite{Smit:2004kh} that CP violation needs not be parametrically small in
the Standard Model. It is well-known that CP violation, even if allowed in the
Standard Model through the Kobayashi-Maskawa mechanism, is a rather elusive
phenomenon. For quarks or Dirac neutrinos it requires the participation of at
least three generations to have a non vanishing result. This is best
summarized by the Jarlskog determinant which involves the CKM matrix through a
very specific combination, the Jarlskog invariant \cite{Jarlskog:1985ht}, and
the quarks masses also in a very specific combination, $\prod_{i<j=u,c,t}
(m_i^2-m_j^2) \prod_{i<j=d,s,b} (m_i^2-m_j^2)$. The Jarlskog determinant is a
twelfth degree polynomial in the masses which must be present, as a factor, in
any CP violating contribution \cite{Farrar:1993sp}. If the Jarlskog
determinant is simply compensated with the appropriate power of $v$, the Higgs
vacuum expectation value, one obtains extremely small ratios: the
dimensionless ratio obtained by dividing by $v^{12}$ gives a number as small
as $10^{-24}$ for quarks. This is a parametrically small result that comes
from assuming a perturbative treatment for the fermion masses. The observation
in \cite{Smit:2004kh} is that, instead of polynomials one should expect
rational functions (plus logarithms) of the fermion masses and this may lead
to a substantial increase in the estimate of the strength of the couplings to
CP violating operators at zero temperature (the only case we consider
throughout this work).

Calculations along these lines were carried out for quarks and dimension six
operators in \cite{Hernandez:2008db} for the P odd sector only, and in
\cite{GarciaRecio:2009zp} for the two parity sectors and including also the
Higgs field.  Unfortunately, the results of the two groups, obtained by two
different methods, are mutually incompatible. The result obtained in
\cite{GarciaRecio:2009zp}, has been reproduced in \cite{Brauner:2011vb} using
the same method as in \cite{GarciaRecio:2009zp} and also in
\cite{Salcedo:2011hy}, this time using the same method as in
\cite{Hernandez:2008db}.  The couplings to selected dimension eight operators
have been obtained in \cite{Salcedo:2011hy} and \cite{Brauner:2012gu}. For
Dirac particles these are the first instances of P odd CP violating
contributions. Extensions to finite temperature have been addressed in
\cite{Brauner:2011vb,Brauner:2012gu}.

In those calculations one indeed finds a large enhancement in the value of the
couplings, as compared to perturbative estimates. Such larger couplings would
have an impact on the viability of cold electroweak baryogenesis scenarios
\cite{Tranberg:2009de,Konstandin:2011ds,Brauner:2011vb}.  Ultimately, the
enhancement comes from the fact that the typical scale in the coupling is not
set by value of the Higgs condensate but rather by the quark masses themselves
and some of them are relatively small.\footnote{Equivalently, disregarding the
  very disparate scales in the values of the Yukawa couplings is not a good
  enough estimate.} However, the precise combinations of masses are not
obvious without a detailed calculation. For dimension six operators, what is
actually found is that the coupling comes from a loop momentum integral which
would be afflicted by infrared (IR) divergences for massless $u$, $d$ and $s$
quarks. As a consequence, finite but different results are obtained depending
on how the ratios between light quark masses are taken in that massless limit.

The coupling to dimension six CP violating operators just discussed does not
have a contribution from leptons in the strict Standard Model, where neutrinos
are massless. In fact, the leptonic loop exactly preserves CP symmetry for
massless neutrinos.  Nevertheless, the scheme used for quark applies quite
directly to massive neutrinos of Dirac type. The small neutrino masses calls
for an investigation of how the possible IR divergencies affect the couplings
in the leptonic sector. In some sense the leptonic case is cleaner than the
quarkonic one, as gluonic corrections (which start at dimension 8) are not
present. On the other hand, the information on neutrino masses and the PMNS
matrix is currently less complete than for quarks.  In addition, neutrinos may
have mass terms of Majorana type that can be accommodated in the Standard
Model invoking a seesaw mechanism
\cite{Yanagida:1980xy,GellMann:1980vs,Morii:2004tp,Giunti:2007ry}. It is of
interest to investigate how the small masses of Majorana type reflect on the
couplings to CP violating operators. This requires a full new determination
of the couplings, as the Dirac results can not be directly adapted to describe
the Majorana case. In this work we consider these two cases, pure Dirac and
pure Majorana neutrinos, with three light flavors although some of the
formulas are more general. As we show, the induced CP violating operators have
at least dimension six. The mixed case, with mass terms of Dirac and Majorana
type simultaneously, is also interesting as it allows dimension four CP odd
operators but it is beyond the scope of the present work.

Section \ref{sec:2} reviews aspects of chiral gauge fermions with Dirac mass
terms. There we present a new derivation of the technique first introduced in
\cite{Salcedo:2008tc} to reduce normal and abnormal parity components of the
fermionic effective action to a gauge covariant Klein-Gordon approach, based
on the operator $K$. In the second part of that section we adapt the previous
approach to include Majorana mass terms, in addition to the Dirac mass ones,
in such a way that the effective action also follows from the determinant of
$K$.

In Section \ref{sec:3} we spell out how the the previous formalism applies to
the leptonic sector of the Standard Model. Before restricting ourselves to the
cases of pure Dirac or pure Majorana neutrino masses, in the second part of
the section we briefly discuss the general case of mixed Dirac plus Majorana
masses.

Section \ref{sec:4} discusses the extraction of the CP odd component of the
effective action with an analysis on the types of allowed contributions. There
it is shown that also for Majorana neutrinos the leading CP violating terms
are of dimension six, with four $W$ fields. However, a new lepton-number
violating mechanism is present in the Majorana case, which works even for two
generations, in addition to the usual Kobayashi-Maskawa mechanism already
present in the Dirac neutrino or quark cases. New mechanisms are found in the
mixed Dirac-Majorana case which involves no charged gauge bosons.

The operator $K$ for the Standard Model with Dirac or Majorana neutrinos is
constructed in detail in Section \ref{sec:5}. A direct application of the
definition of $K$ in the Majorana case leads to expressions with inverse
powers of the neutrino mass matrix, although they dissappear from the final
amplitudes. In that section we show how to remove these inverse powers
directly from the $K$ matrix, from which the effective action follows.

Section \ref{sec:6} describes the explicit computation of the effective action
in the CP odd sector, for the lowest dimensional operators. The calculation is
based on the technique of covariant symbols
\cite{Pletnev:1998yu,Salcedo:2006pv} which directly delivers covariant
operators in the derivative expansion. In that section the allowed operators
are listed together with relations among them from integration by parts,
Bianchi identities or transference of Lorentz indices from ``metric'' type to
``exterior algebra'' type. Explicit results for the couplings are given in
terms of momentum integrals involving masses and the PMNS matrix. Analytical
regularities in the results are discussed there.

Section \ref{sec:7} is devoted to analyzing the results obtained in the previous
section. The new invariants that emerge in the Majorana case, in addition to
the usual Jarlskog invariant are identified. Taking advantage of the small
neutrino masses, reliable approximate formulas are derived for the couplings
of Dirac type and Majorana type. The formulas are particularized for three
typical scenarios considered in the literature, namely, quasi degenerate
neutrino masses, and normal and inverted hierarchies. Numerical estimates for
the couplings to CP violating operators are given for each of the scenarios.
For Dirac neutrinos, it is shown that the different light-heavy patterns in
the quark and lepton sectors, as regards to weak isospin, imply a suppression
of lepton contribution as compared to quarks contributions.  At the same time,
for Majorana neutrinos new operators are activated at leading order in the P
violating sector.

Section \ref{sec:8} summarizes our conclusions.

\section{\textsf{Dirac and Majorana chiral gauge fermions}}
\label{sec:2}

\subsection{\textsf{Dirac fermions}}

We start by reviewing Dirac chiral gauge fermions since eventually Majorana
fermions will be reduced to this case. We closely follow the exposition in
\cite{GarciaRecio:2009zp} and use the same conventions, so further details can
be looked up in that reference. For convenience we work in Euclidean space.
The rules to go back and forth between Minkowskian and Euclidean spaces can be
found in \cite{GarciaRecio:2009zp}.

For Dirac fermions we consider a generic Lagrangian of the form
\begin{equation}\begin{split}
\L(x) &= \bar{\psi}(x)\D\psi(x)
\\ &= \bar{\psi}_R \sD_R\psi_R + \bar{\psi}_L \sD_L\psi_L + 
\bar{\psi}_L m_{LR}\psi_R + \bar{\psi}_R m_{RL}\psi_L
\end{split}
\label{eq:2.1}
\end{equation}
where
\begin{equation}
D^{L,R}_\alpha= \partial_\alpha + V^{L,R}_\alpha
\end{equation}
and $V^{L,R}_\alpha(x)$ and $m_{LR}(x)$ and $m_{RL}(x)$ are external bosonic
fields which are matrices in the internal space of the fermions. (Euclidean)
unitarity requires
\begin{equation}
\mlr(x)=\mrl^\dagger(x) ,\quad
V_{L,R}^\dagger(x) = -V_{L,R}(x)
.
\label{eq:2.3}
\end{equation}
In the chiral representation of the Dirac gammas, $\psi_R$ and $\bar\psi_L$
have only upper components, and $\psi_L$ and $\bar\psi_R$ have only lower
components. The fermionic sector of the Standard Model fits in the scheme of
\Eq{2.1} when all fermions are of Dirac type \cite{GarciaRecio:2009zp}. Later
below we show that it also can accommodate Majorana fermions.

Integration of the fermionic fields provides the effective action
\begin{equation}\begin{split}
& Z = \int D\psi_L D\psi_R D\bar{\psi}_L D\bar{\psi}_R\, e^{-\int d^4x \L(x) }
 = \Det\, \D
,\\
& \Gamma[\mlr,\mrl,V_L,V_R] = -\log Z = -\Tr\,\log \D
.
\end{split}\end{equation}
This functional just sums all one-loop Feynman graphs with the fermion running
on the loop with bosonic external fields attached to it. In this paper by
\emph{effective action} we will always mean the \emph{one-loop} effective
action from integration of the fermions, and not the full effective action
which would include higher loop graphs with internal gauge and Higgs boson
lines.

The effective action is invariant under CP transformations
\begin{equation}
\begin{split}
& \mlr(x) \to \mlr^*(\tilde x),\quad
  \mrl(x) \to \mrl^*(\tilde x),\quad
\\
& V_{R,\alpha}(x) \to \pi_{\alpha\beta} V^*_{R,\beta}(\tilde x),\quad
  V_{L,\alpha}(x) \to \pi_{\alpha\beta} V^*_{L,\beta}(\tilde x),
\end{split}
\label{eq:2.5}
\end{equation}
with $\pi_{\alpha\beta}=\text{\rm diag}(1,-1,-1,-1)$, ~${\tilde x}_\alpha= \pi_{\alpha\beta}x_\beta$.

The effective action can be naturally separated into its parity preserving and
parity violating components,
\begin{equation}
\Gamma = \Gamma^+ + \Gamma^-
.
\end{equation}
$\Gamma^-$ and $\Gamma^+$ are the components with and without the Levi-Civita
pseudotensor, respectively. As a consequence of CPT invariance, $\Gamma^+$ is
purely real and $\Gamma^-$ is purely imaginary, in Euclidean space
\cite{AlvarezGaume:1983ig}. Therefore, modulo ultraviolet (UV) ambiguities,
\begin{equation}
\Gamma^\pm = -\frac{1}{2}(\Tr\,\log\D \pm \Tr\,\log\D^\dagger)
.
\label{eq:2.7}
\end{equation}

The Lagrangian $\L(x)$ is invariant under local chiral transformations. To
expose the chiral properties it will prove convenient to write the Lagrangian
in matricial form, namely,
\begin{equation}
\L(x) = \begin{pmatrix}\bar\psi_L &\bar\psi_R  \end{pmatrix}
\begin{pmatrix}
\mlr & \sD_L \\ \sD_R & \mrl 
\end{pmatrix}
\begin{pmatrix}
\psi_R \\ \psi_L
\end{pmatrix}
.
\label{eq:2.8}
\end{equation}
Chiral gauge transformations take the form
\begin{equation}
 \D \to \D^\Omega = 
\begin{pmatrix}
\Omega^\dagger_L(x) & 0 \\ 0 & \Omega^\dagger_R(x)
\end{pmatrix}
\D
\begin{pmatrix}
\Omega_R(x) &  0 \\ 0 & \Omega_L(x)
\end{pmatrix}
\end{equation}
where $\Omega_{L,R}(x)$ are unitary matrices in internal space.

When $\Omega_L=\Omega_R$ (vector transformations) $\D$ y $\D^\Omega$ are
related by a similarity transformation, as a consequence they have the same
spectrum and the same effective action. In the general chiral case the two
effective actions $\Gamma(\D)$ and $\Gamma(\D^\Omega)$ have equal \emph{UV
  convergent} contributions, since these are unambiguously fixed by the
Lagrangian, but may differ in \emph{UV divergent} ones. More specifically,
from \Eq{2.7} it follows that $\Gamma^+$ can be obtained from the determinant
of $\D\D^\dagger$. This latter operator transforms under a similarity
transformation,
\begin{equation}
 \D \D^\dagger \to 
\begin{pmatrix}
\Omega^\dagger_L(x) &  \\  & \Omega^\dagger_R(x)
\end{pmatrix}
\D\D^\dagger
\begin{pmatrix}
\Omega_L(x) &  \\  & \Omega_R(x)
\end{pmatrix}
,
\end{equation}
and so it can be regularized in a chirally invariant manner. On the contrary
$\Gamma^-$ has a chiral variation, the chiral anomaly, which cannot be
consistently removed \cite{Adler:1969gk,Bell:1969ts,Bardeen:1969md} The chiral
anomaly is saturated by the gauged Wess-Zumino-Witten action (WZW) action
\cite{Wess:1971yu,Witten:1983tw}, and the remaining terms in $\Gamma^-$ are
chirally invariant. Denoting by $\Gamma_c$ the chirally invariant component of
$\Gamma$, one has
\begin{equation}
\Gamma^+ = \Gamma_c^+ ,\quad
\Gamma^- = \Gamma_c^- + \Gamma_\text{\rm gWZW} 
.
\label{eq:2.11}
\end{equation}

The anomalous gauged WZW is known in closed-form
and, as we will argue below, it gives no contribution to CP violation, with
either Dirac or Majorana neutrinos.

The chiral invariant reminder $\Gamma_c[m,V]$ is a functional of the external
fields that admits no closed-form in general. Therefore expansions, such as a
the derivative expansion, must be adopted. Nevertheless the chiral invariance
of $\Gamma_c$ implies a large simplification in the calculations since
everything can be expressed in terms of $\mlr$ and $\mrl$, their chiral
covariant derivatives, and the field strengths,
\begin{equation}
\begin{split}
\hat{D}_\alpha\mlr &= D^L_\alpha\mlr - \mlr D^R_\alpha,
\qquad 
\hat{D}_\alpha\mrl = D^R_\alpha\mrl - \mrl D^L_\alpha,
\\
F^{L,R}_{\alpha\beta} &= [D^{L,R}_\alpha,D^{L,R}_\beta]
.
\end{split}
\label{eq:2.12}
\end{equation}

A complete calculation of $\Gamma_c$ to four covariant derivatives can be
found in \cite{Salcedo:2000hp,Salcedo:2000hx,Salcedo:2008bs}.  However, the
calculation gets very involved for higher orders. As we argue below, the sixth
order is needed in the derivative expansion to pick up the leading CP
violating terms of the effective action.\footnote{At least for pure Dirac or
Majorana neutrinos (see \Eq{4.11}).} To obtain those we follow
\cite{GarciaRecio:2009zp} and take the approach of \cite{Salcedo:2008tc},
although here we present an alternative derivation.

Consider the well-known relation \cite{Negele:1988vy}
\begin{equation}
\int d^n\psi d^n\bar\psi\, e^{-\bar\psi M \psi + \bar\eta \psi + \bar\psi \eta}
= \det\, M \, e^{\bar\eta M^{-1} \eta}
,
\end{equation}
and separate the Grassman variables in two types
\begin{equation}
\bar\psi M \psi = 
\begin{pmatrix}
\bar\psi_1 & \bar\psi_2 
\end{pmatrix}
\begin{pmatrix}
M_{1,1} & M_{1,2} \\ M_{2,1} & M_{2,2}
\end{pmatrix}
\begin{pmatrix}
\psi_1 \\ \psi_2 
\end{pmatrix}
\end{equation}
where the $M_{i,j}$ are themselves matrices in general. By integrating
first $\psi_1$ and $\bar\psi_1$, and then $\psi_2$ and $\bar\psi_2$, or the
other way around, the following identities are obtained
\begin{equation}
\begin{split}
\det\, M &= \det\, M_{2,2} \, \det\left(M_{1,1} - M_{1,2}M^{-1}_{2,2} M_{2,1}\right)
\\
&= \det\, M_{1,1} \, \det\left(M_{2,2} - M_{2,1}M^{-1}_{1,1} M_{1,2}\right)
.
\end{split}
\label{eq:2.15}
\end{equation}
These identities can be applied directly to the chiral fermions in \Eq{2.8}:
\begin{equation}
\Det\,\D = \Det(\kappa_L) = \Det(\bar\kappa_R)
\end{equation}
with\footnote{Here we are assuming that $\mlr$ is a square and regular
  matrix. This is natural in parity preserving theories but not in chiral
  theories. Nevertheless this case is sufficiently general for our purposes.}
\begin{equation}
\begin{split}
\kappa_L &= \mlr \mrl - \sigma_\alpha D_L{}_\alpha \mrl^{-1} \bar\sigma_\beta D_R{}_\beta \mrl
\,,
\\
\bar\kappa_R &= \mrl \mlr - \bar\sigma_\alpha D_R{}_\alpha \mlr^{-1} \sigma_\beta
D_L{}_\beta \mlr
.
\end{split}
\end{equation}
Here, $\sigma_\alpha$ and $\bar\sigma_\alpha$ are the Pauli and identity matrices
corresponding to the chiral representation of the the Dirac gammas:
\begin{equation}
\gamma_\alpha = 
\begin{pmatrix}
0 & \sigma_\alpha \\ \bar\sigma_\alpha & 0
\end{pmatrix}
,\quad
\gamma_5 = 
\begin{pmatrix}
1 & 0 \\ 0 & -1
\end{pmatrix}
.
\end{equation}
To avoid working with Dirac bispinors, we introduce the operators
\begin{equation}
\begin{split}
K_L &= \mlr \mrl - \sD_L \mrl^{-1} \sD_R \mrl 
 = \begin{pmatrix}
\kappa_L & 0 \\ 0 & \bar{\kappa}_L
\end{pmatrix}
, \\
K_R &= \mrl \mlr - \sD_R \mlr^{-1} \sD_L \mlr 
 = \begin{pmatrix}
\kappa_R &  0 \\ 0 & \bar{\kappa}_R
\end{pmatrix}
.
\end{split}
\label{eq:2.19}
\end{equation}
These two operators are related through the identity 
\begin{equation}
K_R = \mlr^{-1} K_L^\dagger
\mlr
.
\label{eq:2.19a}
\end{equation}
In terms of these operators
\begin{equation}
\begin{split}
\Tr(\log \kappa_L) &= \Tr(P_R \log \,K_L),\qquad
\Tr(\log \bar\kappa_R) = \Tr(P_L \log \,K_R)
,
\\
\quad
P_{L,R} &= \frac{1}{2}(1 \mp \gamma_5 )
,
\end{split}
\end{equation}
and we finally obtain
\begin{equation}
\begin{split}
\Gamma_c &= - \Tr(P_R \log \, K_L )  = - \Tr(P_L \log \, K_R )
\,,
\\
\Gamma_c^+ &= -\frac{1}{2}\Tr(\log K_L) = -\frac{1}{2}\Tr(\log K_R)
,
\\
\Gamma_c^- &= -\frac{1}{2}\Tr(\gamma_5 \log K_L) 
= +\frac{1}{2}\Tr(\gamma_5\,\log K_R)
.
\label{eq:2.20}
\end{split}
\end{equation}
To make the identifications with $\Gamma_c^+$ and $\Gamma_c^-$ above, we have
used that only $\gamma_5$ introduces the Levi-Civita pseudotensor after taking
Dirac traces. On the
other hand the identity \neq{2.19a} implies that
$\Gamma_c^+$ is purely real and $\Gamma_c^-$ is purely imaginary (in Euclidean
space).

The usefulness of the relations (\ref{eq:2.20}) is two-fold, first, $K_{L,R}$
are operators of Klein-Gordon type. While this was already the case for
$\D\D^\dagger$, to obtain $\Gamma_c^+$, no such operator was available for
$\Gamma_c^-$ before \cite{Salcedo:2008tc}. This allows a substantial
simplification of the Dirac algebra in the calculations. And second, $K_{L,R}$
are chiral covariant, that is,
\begin{equation}
K_L \to \Omega_L^\dagger K_L\Omega_L,
\quad
K_R \to \Omega_R^\dagger K_R\Omega_R
.
\end{equation}
This fact guarantees that explicit chiral gauge invariance can be maintained
throughout the calculation, also in the parity odd component of the effective
action (the component afflicted by chiral anomalies).

It is also of interest to see the relation of the operators $K_{L,R}$ with an
{\em effective Lagrangian} of the Klein-Gordon type for the fermionic
amplitudes. The propagator is given by
\begin{equation}
\left\langle 
\begin{pmatrix} \psi_R & \psi_L \end{pmatrix}(x)
\begin{pmatrix} \bar{\psi}_L \\ \bar{\psi}_R \end{pmatrix}(x^\prime)
\right\rangle
=
\langle x | \D^{-1} | x^\prime \rangle
.
\end{equation}
On the other hand, the inverse Dirac operator can be written as
\begin{equation}
\D^{-1} =
\begin{pmatrix}
( \mlr - \sD_L \mrl^{-1} \sD_R )^{-1}
&
(\sD_L - \mlr \sD_R^{-1} \mrl )^{-1}
\\
(\sD_R - \mrl \sD_L^{-1} \mlr )^{-1}
&
( \mrl - \sD_R \mlr^{-1} \sD_L )^{-1}
\end{pmatrix}
.
\end{equation}
Comparing with the definitions of $K_{L,R}$ in \Eq{2.19}, it follows that
\begin{equation}
\langle x | P_R K_L^{-1} | x^\prime \rangle
= \mrl^{-1}\langle \psi_R(x) \bar{\psi}_L (x^\prime)\rangle 
,\quad
\langle x | P_L K_R^{-1} | x^\prime \rangle =
 \mlr^{-1} \langle \psi_L (x)\bar{\psi}_R (x^\prime)\rangle
.
\label{eq:2.24}
\end{equation}
Therefore the effective Lagrangian $\bar{\psi}_L K_L \tilde{\psi}_R(x)$, with
$\tilde{\psi}_R \equiv \mrl^{-1}\psi_R$, correctly describes the propagator
$\langle \psi_R \bar{\psi}_L\rangle$, although it gives no direct information
on, e.g., $\langle \psi_L \bar{\psi}_L\rangle$, and similarly for
$K_R$.

The manipulations leading to (\ref{eq:2.15}) and (\ref{eq:2.20}) are based on
the formal identity 
\begin{equation}
\Tr\log(AB)=\Tr\log(A)+\Tr\log(B)
,
\end{equation}
where $A$ and $B$ are differential or pseudodifferential operators. In fact,
the determinant of these operators contains UV divergences which have to be
removed by means of some renormalization procedure (e.g., $\zeta$-function
\cite{Dowker:1975tf,Hawking:1976ja,Elizalde:1994gf}). The choice of
renormalization introduces finite ambiguities which can give corrections to
the formal identity. This is the origin of the quantum anomalies and the
gauged WZW term above. (For a careful treatment including
everything as in \neq{2.11} see \cite{Ball:1988xg,Salcedo:2000hx,Salcedo:2008tc}.) On the
other hand, the formal identity holds for the UV finite contributions. Within
the derivative expansion, the UV convergent terms are those of order six and
higher which we will compute below for the CP odd sector. Therefore no quantum
ambiguities nor anomalies appear in the CP violating sector.

\subsection{\textsf{Majorana fermions}}

The fermion Lagrangian including Majorana mass terms is of the
form\footnote{Further ``Majorana'' vector couplings can be considered, of the
  type $\displaystyle \frac{1}{2}\bar{\psi}_L^c \slashchar{A}_{LR} \psi_R +
  \frac{1}{2}\bar{\psi}_R^c \slashchar{A}_{RL} \psi_L + \text{H.c.}$, however,
  such terms are not present in the Standard Model or any renormalizable
  theory. Renormalizability would require these vector fields to be of gauge
  type and the gauge group would mix different chiralities which is forbidden
  for an internal symmetry.}
\begin{equation}\begin{split}
\L(x) =& \bar{\psi}_R \sD_R\psi_R + \bar{\psi}_L \sD_L\psi_L + 
\bar{\psi}_L m_{LR}\psi_R + \bar{\psi}_R m_{RL}\psi_L
\\ 
&
+ \frac{1}{2}\bar{\psi}_L^c m_L \psi_L 
+ \frac{1}{2}\bar{\psi}_L m_L^\dagger \psi_L^c
+ \frac{1}{2}\bar{\psi}_R^c m_R \psi_R 
+ \frac{1}{2}\bar{\psi}_R m_R^\dagger \psi_R^c
\,.
\end{split}
\label{eq:2.1b}
\end{equation}
Here, as usual, $\psi_{L,R}^c \equiv C\bar{\psi}^T_{L,R}$ and
$\bar{\psi}_{L,R}^c \equiv - \psi^T_{L,R} C^\dagger$, $C$ being the unitary
matrix such that $C^\dagger\gamma_\alpha C =-\gamma_\alpha^T$. The Majorana mass
complex matrices $m_L(x)$ and $m_R(x)$ are symmetric since $C$ is
antisymmetric. Under CP and chiral gauge transformations they transform,
respectively, as
\begin{equation}
\begin{split}
&
m_L(x) \to m_L^\dagger(\tilde x)
,\quad
m_R(x) \to m_R^\dagger(\tilde x)
,
\\
&
m_L(x) \to \Omega_L^T(x)m_L(x) \Omega_L(x)
,\quad
m_R(x) \to \Omega_R^T(x)m_R(x) \Omega_R(x)
.
\end{split}
\label{eq:2.26a}
\end{equation}

In the Euclidean formulation, $\psi_{L,R}(x)$ and $\bar{\psi}_{L,R}(x)$ are
independent fields to be integrated over. On the other hand, $\psi_{L,R}^c$,
$\bar{\psi}_{L,R}^c$ are merely auxiliary variables ($\bar\psi_L^c$ has the
same content as $\psi_L$, etc). The Lagrangian with Dirac and Majorana mass
terms can be written in matrix form using the trick of \emph{duplicating} the
size of the matrices, specifically (cf. \Eq{2.8})
\begin{equation}
\L(x) = \frac{1}{2}
\begin{pmatrix}
\psi_R^T & \psi_L^T & \bar{\psi}_R &  \bar{\psi}_L 
\end{pmatrix}
\begin{pmatrix}
-C^\dagger m_R & 0 & -\sD_R^T & -\mlr^T \\
0 & -C^\dagger m_L & -\mrl^T & -\sD_L^T \\
\sD_R & \mrl & C m^\dagger_R & 0 \\
\mlr & \sD_L & 0 & Cm_L^\dagger 
\end{pmatrix}
\begin{pmatrix}
\psi_R \\ \psi_L \\ \bar{\psi}_R^T \\  \bar{\psi}_L^T 
\end{pmatrix}
.
\label{eq:2.25}
\end{equation}
As it should be, the operator in between is antisymmetric, the fermion fields
being Grassmann c-numbers. Functional integration on $\psi_{L,R}(x)$ and
$\bar{\psi}_{L,R}(x)$ gives the Pffaffian of this operator, which equals the
square root of its determinant. Having reduced the problem to a determinant,
one can apply similarity transformations to remove $C$ to enforce explicit
Lorentz invariance, as well as rearrangement of rows and of columns to obtain
a suitable form. Specifically, we take
\begin{equation}
\L(x) = \frac{1}{2}
\begin{pmatrix}
\bar\psi_L & \bar\psi_R^c & \bar{\psi}_R & \bar{\psi}_L^c 
\end{pmatrix}
\begin{pmatrix}
\mlr & m_L^\dagger & \sD_L & 0 \\
m_R & \mlr^T & 0 & \sD_R^* \\
\sD_R & 0 & \mrl & m_R^\dagger \\
0 & \sD_L^* & m_L & \mrl^T
\end{pmatrix}
\begin{pmatrix}
\psi_R \\ \psi_L^c \\ \psi_L \\ \psi_R^c 
\end{pmatrix}
:=\frac{1}{2}\bar\Psi  \D_2 \, \Psi
\,.
\label{eq:2.26}
\end{equation}
It is important to remark that in \Eq{2.25} $\sD^T=\gamma_\alpha^T D_\alpha^T$,
while in \Eq{2.26} we use the notation $\sD^*:=\gamma_\alpha D_\alpha^*$ (no complex
conjugation on the Dirac gammas). \emph{We will adopt the same notation in
  what follows.} Also, $D^*_\alpha=\partial_\alpha + V^*_\alpha$ and
$D^T_\alpha=-\partial_\alpha+V^T_\alpha= -D^*_\alpha$.

As already noted the partition function equals $(\Det\,\D_2)^{1/2}$, therefore
\begin{equation}
\Gamma[\mlr,\mrl,m_L,m_R,V_L,V_R] = -\frac{1}{2}\Tr\,\log \D_2
.
\label{eq:2.27}
\end{equation}

The order of the fields in $\D_2$ has been chosen so that the problem of
computing $\Det\,\D_2$ is identical to that for pure Dirac fermions in
\Eq{2.8}, with the replacements
\begin{equation}
\begin{split}
&
\psi_{L,R} \to \begin{pmatrix} \psi_{L,R} \\ \psi_{R,L}^c \end{pmatrix}
,\quad
\bar{\psi}_{L,R} \to \begin{pmatrix} \bar{\psi}_{L,R} &
  \bar{\psi}_{R,L}^c \end{pmatrix}
,
\\
&
\mlr \to \begin{pmatrix}
\mlr & m_L^\dagger  \\
m_R & \mlr^T 
\end{pmatrix}
, \quad
\mrl \to 
\begin{pmatrix}
 \mrl & m_R^\dagger \\
 m_L & \mrl^T
\end{pmatrix}
, \quad
\\
&
D_L{}_\alpha \to \begin{pmatrix}
 D_L{}_\alpha & 0 \\
 0 & D_R^*{}_\alpha 
\end{pmatrix}
, \quad
D_R{}_\alpha \to \begin{pmatrix}
 D_R{}_\alpha & 0 \\
 0 & D_L^*{}_\alpha 
\end{pmatrix}
.
\end{split}
\end{equation}
In particular, the explicit results in
\cite{Salcedo:2000hp,Salcedo:2000hx,Salcedo:2008bs} for the effective action
of Dirac fermions can be immediately extended to Majorana fermions using the
above identifications.

The chiral transformation of $\D_2$ is given by
\begin{equation}
\D_2 \to 
\begin{pmatrix}
\Omega^\dagger_L & & & \\  & \Omega^T_R & & \\
& & \Omega^\dagger_R &  \\ & & & \Omega^T_L
\end{pmatrix}
\D_2
\begin{pmatrix}
\Omega_R & & & \\  & \Omega^*_L & & \\
& & \Omega_L &  \\ & & & \Omega^*_R
\end{pmatrix}
,
\end{equation}
so the chiral invariant part of the effective action will involve the chiral
covariant pieces in \Eq{2.12}, plus the new chiral covariant derivatives
\begin{equation}
\hat{D}_\alpha m_L = D^{L *}_\alpha m_L - m_L D^L_\alpha
,\quad
\hat{D}_\alpha m_R = D^{R *}_\alpha m_R - m_R D^R_\alpha
.
\label{eq:2.31}
\end{equation}

\section{\textsf{The (extended) Standard Model}}
\label{sec:3}

\subsection{\textsf{Leptonic sector of the Standard Model}}

The fermionic sector of the Standard Model of particle physics, extended to
include either Dirac, Majorana or mixed neutrino masses can be
accommodated in the scheme of the previous section. First we consider
explicitly the general case where both Dirac and Majorana neutrino masses are
present and eventually we will restrict ourselves to the simpler cases of pure
Dirac or pure Majorana neutrino masses.

In order to apply \Eq{2.26} to the leptonic sector of the extended Standard
Model we take the identifications
\begin{equation}
\bar{\psi}_{L,R} = \begin{pmatrix}\bar{\nu}_{L,R} & \bar{e}_{L,R}\end{pmatrix} 
 , \quad
\psi_{L,R} = \begin{pmatrix}\nu_{L,R} \\ e_{L,R}\end{pmatrix}
,
\label{eq:3.1}
\end{equation}
where $e_{L,R}(x)$ is the field of the charged leptons. This is a Dirac spinor
as well as a vector on generation or family space, containing the electron,
muon and tau fields. Likewise, the vector $\nu_{L,R}(x)$ represents the fields
of the three left-handed neutrinos, and a certain number $N_s$ of right-handed
ones. The dimensions of $\nu_L$, $\nu_R$, $e_L$ and $e_R$ are $g$, $N_s$, $g$
and $g$, respectively, where $g=3$ is the number of generations.

Further, for the mass terms in \Eq{2.26}
\begin{equation}
\mlr = \mrl^\dagger = \begin{pmatrix}
\frac{\phi}{v} M_D & 0 \\ 0 & \frac{\phi}{v} M_e
 \end{pmatrix} 
,\quad
m_L = \begin{pmatrix}
\frac{\phi^2}{v^2} M_L & 0 \\ 0 & 0
 \end{pmatrix}
,\quad
m_R = \begin{pmatrix}
M_R & 0 \\ 0 & 0
 \end{pmatrix}
.
\label{eq:3.2}
\end{equation}
$M_D$ and $M_e$ are constant complex matrices in generation space representing
the Dirac mass matrices of neutrinos and charged leptons,
respectively. Similarly $M_L$ and $M_R$ are the Majorana mass matrices of left
and right handed neutrinos, respectively. The dimensions of $M_D$, $M_e$,
$M_L$ and $M_R$ are $g \times N_s$, $g \times g$, $g \times g$ and $N_s \times
N_s$, respectively. 

We adopt the unitary gauge throughout. $\phi(x)$ is the neutral Higgs field in
that gauge and $v$ its vacuum expectation value. The coupling of the mass
terms to the Higgs field adopted here takes into account that
$\bar{\nu}_R\nu_L$ and $\bar{e}_R e_L$ are $\SU(2)$ doublets, as the Higgs,
while $\bar{\nu}_L^c\nu_L$ is a triplet and $\bar{\nu}_R^c\nu_R$ is a singlet
\cite{Morii:2004tp}.

Finally, the covariant derivatives in \Eq{2.26}, take the following form in
the Standard Model:
\begin{equation}
D^L_\alpha = \begin{pmatrix}
D^\nu_\alpha + Z_\alpha & W^+_\alpha \\
W^-_\alpha & D^e_\alpha - Z_\alpha 
 \end{pmatrix}
, \quad
D^R_\alpha = \begin{pmatrix}
D^\nu_\alpha & 0 \\
0 & D^e_\alpha
 \end{pmatrix}
.
\label{eq:3.3}
\end{equation}
$W_\alpha^{\pm}$ represent the fields of the charged bosons and $Z_\alpha$ the
field of the $Z^0$. For convenience we have included the $\SU(2)\otimes \U(1)$
couplings in the gauge fields. The relation to the canonically normalized
fields (denoted with tilde) is as follows
\cite{GarciaRecio:2009zp,Huang:1992bk}
\begin{equation}
W_\alpha^\pm = \frac{1}{\sqrt{2}}g \tilde{W}^\pm_\alpha
,\quad
Z_\alpha 
=
\frac{1}{2}\frac{g}{\cos\theta_W}\tilde{Z}_\alpha
= \frac{1}{2}g\tilde{W}^3_\alpha
-\frac{1}{2}g^\prime\tilde{B}_\alpha
 ,
\end{equation}
where $\tilde{B}_\alpha$ is the gauge field of the weak hypercharge $\U(1)$ group
and $\theta_W$ the weak angle. On the other hand, $D_\nu$ and $D_e$ are
covariant derivatives, corresponding to the remaining gauge freedom within the
unitary gauge. Specifically,
\begin{equation}
\begin{split}
D^\nu_\alpha &= \partial_\alpha, \quad  A^\nu_\alpha = 0,
\\
D^e_\alpha &= \partial_\alpha + A^e_\alpha
,\quad
A^e_\alpha = -g^\prime \tilde{B}_\alpha
=
-e \tilde{A}_\alpha +2 \sin^2\theta_W \, Z_\alpha
,
\end{split}
\label{eq:3.5}
\end{equation}
where $\tilde{A}_\alpha$ is the photon field and $-e$ the electron electric
charge. In the following we will work with $Z_\alpha$ and $D^e_\alpha$ as basic
variables, but it should be remembered that $Z_\alpha$ also appears in
$D^e_\alpha$ when the final results are expressed in terms of the physical
fields $\tilde{Z}_\alpha$ and $\tilde{A}_\alpha$.

From the general formula in \Eq{2.12} the following field strengths can be
constructed:
\begin{equation}
\begin{split}
W^+_{\alpha\beta} & := D^\nu_\alpha \, W^+_\beta - W^+_\beta D^e_\alpha
= \partial_\alpha W^+_\beta - A^e_\alpha W^+_\beta 
,
\\
W^-_{\alpha\beta} & := D^e_\alpha \, W^-_\beta - W^-_\beta D^\nu_\alpha
= \partial_\alpha W^-_\beta + A^e_\alpha W^-_\beta 
,
\\
F^e_{\alpha\beta} & = [ D^e_\alpha , D^e_\beta ] = \partial_\alpha A^e_\beta -
\partial_\beta A^e_\alpha
,
\\
Z_{\alpha\beta} &:= D^\nu_\alpha Z_\beta - Z_\beta D^\nu_\alpha =
\partial_\alpha Z_\beta
.
\end{split}
\end{equation}
Let us emphasize that the tensors $W^\pm_{\alpha\beta}$ and $Z_{\alpha\beta}$ just
defined \emph{are not} antisymmetric.

\begin{table}[t]
\centering
\begin{tabular}{ccc}
\hline 
Quantity & Complex & CP \cr
& conjugation &
\cr
\hline
$\phi$ & $\phi$ & $\phi$
\cr
$W^\pm_{\alpha}$ & $-W^\mp_{\alpha}$ & $-W^\mp_{\alpha}$
\cr
$Z_{\alpha}$ & $-Z_{\alpha}$ & $-Z_{\alpha}$
\cr
$\varphi_{\alpha}$ & $\varphi_{\alpha}$ & $ \varphi_{\alpha}$
\cr
$F^e_{\alpha\beta}$ & $-F^e_{\alpha\beta}$ & $-F^e_{\alpha\beta}$
\cr
$\epsilon_{\mu\nu\alpha\beta}$ & $\epsilon_{\mu\nu\alpha\beta}$ 
& $-\epsilon_{\mu\nu\alpha\beta}$
\cr
$U$ & $ U^* $ 
& $ U^* $
\cr
\hline
\end{tabular}\\
\caption{Transformation of various quantities (Euclidean version) under
  complex conjugation and CP. The derivatives of the fields follow the same
  rules as the fields themselves.}
\label{tab:1}
\end{table}
The properties of these fields under complex conjugation and CP follow from
Eqs. (\ref{eq:2.3}) and (\ref{eq:2.5}). They are summarized in Table
\ref{tab:1}.

\subsection{\textsf{Neutrinos with mixed Dirac-Majorana mass terms}}

The main topic of the paper is Majorana and Dirac neutrinos. Nevertheless, for
future reference, in the rest of this section we briefly review the case of
general neutrino mass matrices. For simplicity we set $\phi(x)=v$ in this
discussion and assume that there are no accidental mass degeneracies and no
fermion is massless.

Certainly, one can rotate the lepton fields by means of constant unitary
matrices so that the Eqs. (\ref{eq:3.2}) and (\ref{eq:3.3}) are unchanged
except that $M_e$ is replaced by a diagonal and positive matrix $m_e$. Once
this choice is taken, the neutrino mass matrices can be diagonalized by means
of a unitary transformation as follows
\begin{equation}
\begin{pmatrix}
 M_L^* & M_D \\ M_D^T & M_R 
\end{pmatrix}
= \mathcal{U} 
\begin{pmatrix}
m_1 & 0 \\ 0 & m_2 
\end{pmatrix} 
\mathcal{U}^T
,\qquad
\mathcal{U} = \begin{pmatrix}
 A & B \\ D & C 
\end{pmatrix}
,\quad
\mathcal{U}^{-1} = \mathcal{U}^\dagger
,
\label{eq:3.7}
\end{equation}
so that the matrices $m_{1,2}$ are diagonal and positive. $A$ and $C$ are
square submatrices but need not be unitary.

If these rotations are applied to the fermion fields the structure in
\Eq{2.26} is not preserved, since right and left handed fields are
mixed. However the form in \Eq{2.8} still holds, with
\begin{equation}
\begin{split}
\mlr &= \mrl^\dagger = \begin{pmatrix}
0 & 0 & m_1 & 0 \\
0 & m_e & 0 & 0 \\
m_2 & 0 & 0 & 0 \\
0 & 0 & 0 & m_e \\
\end{pmatrix}
,
\\
D^L_\alpha &= \begin{pmatrix}
D^\nu_\alpha + A^\dagger A Z_\alpha & A^\dagger W^+_\alpha &  A^\dagger B Z_\alpha &  0 \\
A W^-_\alpha & D^e_\alpha - Z_\alpha & B W^-_\alpha & 0 \\
B^\dagger A Z_\alpha & B^\dagger W^+_\alpha & D^\nu_\alpha + B^\dagger B Z_\alpha & 0 \\
0 & 0 & 0 & D^{e,*}_\alpha \\
\end{pmatrix}
,
\\
D^R_\alpha &= \begin{pmatrix}
D^\nu_\alpha - B^T B^* Z_\alpha & 0 & -B^T A^* Z_\alpha &  -B^T W^-_\alpha \\
0 & D^e_\alpha & 0 & 0 \\
-A^T B^* Z_\alpha & 0 & D^\nu_\alpha - A^T A^* Z_\alpha & -A^T W^-_\alpha \\
-B^* W^+_\alpha & 0 & -A^* W^+_\alpha & D^{e,*}_\alpha + Z_\alpha \\
\end{pmatrix}
.
\end{split}
\label{eq:3.8}
\end{equation}

Here the fermion fields of \Eq{2.8} correspond to $\psi_R =
(\nu_R,e_R,\nu_L^c,e_L^c)$, $\psi_L=(\nu_L,e_L,\nu_R^c,e_R^c)$, $\bar\psi_R =
(\bar\nu_R, \bar{e}_R, \bar{\nu}_L^c, \bar{e}_L^c)$, and $\bar\psi_L =
(\bar\nu_L, \bar{e}_L, \bar\nu_R^c, \bar{e}_R^c)$. (For simplicity, we denote
rotated and unrotated lepton fields with the same symbols, as we are not using
them in the rest of the paper.)

It is noteworthy that the Dirac operator in \Eq{3.8} is independent of the
submatrices $C$ and $D$ in $\mathcal{U}$. In addition, the determinant of that
Dirac operator, and consequently its effective action, is unchanged if the
matrices $A$ and $B$ are subject to the following transformations:
\begin{equation}
(A,B) \to (V A V_A , V B V_B )
,\quad
V = \text{\rm diag}(e^{\theta_1},\ldots,e^{\theta_g})
,\quad
V_A, V_B = \text{\rm diag}(\pm,\ldots, \pm)
.
\end{equation}
(Here $g$ denotes the number of generations, and the diagonal matrices $V_A$,
and $V_B $ have dimension $g$ and $N_s$, respectively.)

In the limit of large $M_R$ keeping $M_D$ and $M_L$ finite, $B$ vanishes and
$A$ becomes unitary. In this case the neutrinos with mass $m_2$ (denoted
$\nu_R$ above) completely decouple and only left-handed neutrinos remain. This
is equivalent to assuming neutrinos with pure Majorana masses (i.e.,
$M_D=0$) from the beginning. If in addition to large $M_R$, $M_L=0$ is also
assumed, $m_1$ is small, which is the well-known seesaw mechanism to account
for the small masses of the neutrinos
\cite{Yanagida:1980xy,GellMann:1980vs,Morii:2004tp,Giunti:2007ry}. The
coupling to the Higgs field assumed in \Eq{3.2} is also consistent with this
scenario. The coupling between charged and neutral leptons takes the form
$\bar{e}_L A \!\thru{\,W}^- \nu_L$, therefore, in this limit, $A$ is
identified with the PMNS matrix
\cite{Pontecorvo:1957cp,Pontecorvo:1967fh,Maki:1962mu} usually denoted by $U$
in the literature \cite{Beringer:1900zz}.

\section{\textsf{CP odd component of the effective action}}
\label{sec:4}

As said, the effective action $\Gamma[\mlr,\mrl,m_L,m_R,V_L,V_R]$ in \Eq{2.27}
is invariant under a full CP transformation applied to all external fields. In
the Standard Model the physical CP transformation refers to the gauge fields
and Higgs (and to fermions but these are integrated out in the effective
action) while the fermion mass matrices are unchanged. So, to identify the
even and odd components of $\Gamma$ under CP, one can look for the symmetric
and antisymmetric components when the gauge fields and Higgs are
CP-transformed or, more conveniently, when the mass matrices are
CP-transformed. From Eqs. (\ref{eq:2.5}) and (\ref{eq:2.26a}), the latter
amounts to
\begin{equation}
M_L \to M_L^* ,\quad M_R \to M_R^* ,\quad M_D \to M_D^*,\quad M_e \to M_e^* .
\end{equation}
After diagonalization, for Majorana or Dirac neutrinos, the previous
transformation amounts to
\begin{equation}
U \to U^* \qquad \text{(CP transformation)}
.
\end{equation}

For pure Majorana neutrinos ($A = U$ and $B=0$ in \Eq{3.8}) or pure Dirac
neutrinos ($A=0$ and $B=U$), it follows that the presence of $U$ is always
tied to a charged current vertex. Therefore, the CP violating component of
$\Gamma$ must contain $W^\pm$ fields, and in equal number of $W^+$ and $W^-$,
in order to fulfill electric charge conservation, since there are no other
charged external fields. Thus we will consider an expansion of the effective
action in powers of the $W^\pm$ fields:
\begin{equation}
\Gamma = \sum_{n=0}^\infty \Gamma_{2n}
,
\end{equation}
where $\Gamma_{2n}$ contains $n$ $W^+ W^-$ pairs.

\begin{figure}[ht]
\begin{center}
\epsfig{figure=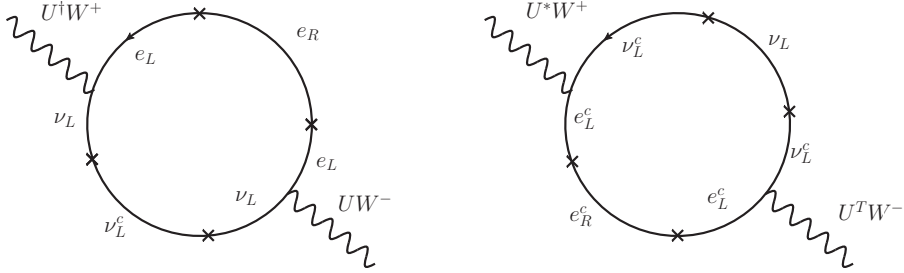,width=120mm}
\caption{Mechanisms present in $\Gamma_2$. Gauge bosons are inwards.}
\label{fig:3}
\end{center}
\end{figure}
From the previous discussion it follows that $\Gamma_0$ cannot have a CP
violating component. The same is true of the $\Gamma_2$. Indeed, within the
formulation on the Dirac operator of the previous section, the two possible
types of Feynman graphs for $\Gamma_2$ are those displayed in
Fig. \ref{fig:3}. It is sufficient to consider one of them since they are
related by conjugation and give the same result. In generation space, a
typical graph has a structure
\begin{equation}
G(U) = \tr ( U^\dagger f_1(m_e) U f_2(m_\nu) ) 
,
\end{equation}
from the vertices $U^\dagger W^+$ and $U W^-$, and $f_1(m_e)$ and
$f_2(m_\nu)$ are \emph{diagonal matrices} from the propagators. Under CP,
\begin{equation}
\begin{split}
G(U) \to G(U^*) &= \tr \left( U^T f_1(m_e) U^* f_2(m_\nu) \right)
= \tr \left( \left(U^T f_1(m_e) U^* f_2(m_\nu) \right)^T \right) 
\\
&=
\tr \left( f_2(m_\nu) U^\dagger f_1(m_e) U  \right)
= G(U)
.
\end{split}
\end{equation}
This result was to be expected: $\Gamma_2$ is structurally identical for
Majorana and Dirac neutrinos, or even for leptons and quarks. As is
well-known, the insertion of just one $W^+$ and one $W^-$ in the quark loop
does not allow the quarks to visit the three generations, which is the minimum
required to have CP violation with Dirac particles \cite{Kobayashi:1973fv}.
Beyond $\Gamma_2$ it is no longer true that the Feynman graphs involving
Majorana and Dirac neutrinos have necessarily the same structure. In the
Majorana case, fermionic number violating terms appear in $\Gamma_4$ that
allow to break CP even for two generations.

Of course, to reach the conclusion that $\Gamma_2$ is CP even it is crucial
that we are considering only the \emph{one-loop} effective action. It is
perfectly possible to write CP violating operators of the type $W^+ W^-$. For
instance
\begin{equation}
Z_\alpha ( W^+_\alpha W^-_{\beta\beta} +  W^+_{\beta\beta} W^-_\alpha )
,
\quad
\epsilon_{\mu\nu\alpha\beta} W^+_{\mu\nu} W^-_{\alpha\beta}
.
\label{eq:5.6}
\end{equation}
The first one is parity even, the second one is parity odd. Our previous
argument implies that such operators require Feynman graphs with internal
gauge boson lines, and this amounts to going beyond one-loop.

We will further use the notation $\Gamma_{2n+d}$ to indicate the component of
the effective action composed of operators with $n$ $W^+ W^-$ pairs and a
total of $2n+d$ Lorentz indices carried by the fields. So the two operators in
\Eq{5.6} are of the type $2+2$. Within a \emph{covariant derivative
  expansion}, $2n+d$ is the order of the operator, that is, the number of
derivatives it carries (in this counting each gauge field or derivative counts
as order 1, the Higgs field is of order 0). Equivalently, $2n+d$ is the
\emph{dimension} of the operator (counting the operator $\phi/v$ as
dimensionless). In an even-dimensional spacetime $d$ is always even.

We have just argued that $\Gamma_{0+d}$ and $\Gamma_{2+d}$ are CP even for any
value of $d$. It is easy to see that the components $\Gamma_{2n+0}$ are also
CP even. Indeed, no CP odd operator can be written using only $W^\pm$ with no
other gauge fields nor derivatives \cite{GarciaRecio:2009zp}.  Since operators
of the type $2n+d>4$ are UV convergent in four dimensions, this implies that
all UV divergent terms of the effective action are CP even. This includes the
gauged WZW term which has dimension four. The first contribution to CP
violation comes from the dimension 6 operators in $\Gamma_{4+2}$ (as said
$\Gamma_{6+0}$ is CP even). These are the operators to be considered in this
work, specifically for Majorana neutrinos. $\Gamma_{4+2}$ for quarks have been
computed in \cite{GarciaRecio:2009zp,Brauner:2011vb}. Some operators of the
type $4+4$ have been calculated for the quark sector in \cite{Salcedo:2011hy}
and of those of the type $6+2$ in \cite{Brauner:2012gu}.

The effective action can be expanded in the form
\begin{equation}
\Gamma = \int d^4x \, \sum_k \left(\frac{v}{\phi(x)}\right)^{d_k-4} 
 g_k \mathcal{O}_k(x)
,
\label{eq:4.7a}
\end{equation}
where the $\mathcal{O}_k$ represent local operators of dimension $d_k=2n+d$,
constructed with gauge fields and their derivatives, as well as derivatives of
the Higgs field. The $g_k$ are the corresponding couplings and they depend on
the lepton mass matrices. The couplings come as integrals over the momentum of
the fermion running in the loop. If underivated $\phi(x)$ are not included in
the operators, they should go in the couplings. However, for Dirac neutrinos,
the dependence on underivated $\phi$ follows from dimensional counting since
the Higgs couples as the fermion masses, and this produces the explicit
dependence shown in (\ref{eq:4.7a}). For Majorana neutrinos, this is no longer
true and the $g_k$, as defined in (\ref{eq:4.7a}), still retain some dependence
on $\phi(x)$ from the neutrino masses.

An operator $\mathcal{O}_k$ is even under charge conjugation if and only it is
hermitian, therefore the CP odd operators are antihermitian in the parity even
sector and hermitian in the parity odd one (Table \ref{tab:1}). Recalling that
$\Gamma^+$ is real and $\Gamma^-$ is imaginary, it follows that \emph{the
  couplings of CP violating operators are purely imaginary}. The same
conclusion follows from noting that in Euclidean space no factor $i$
(imaginary unit) is generated through the Feynman rules, momentum integration
or tracing of Dirac gammas, hence $g_k$ will be imaginary if and only if it is
antisymmetric under $U \to U^*$.

\begin{figure}[ht]
\begin{center}
\epsfig{figure=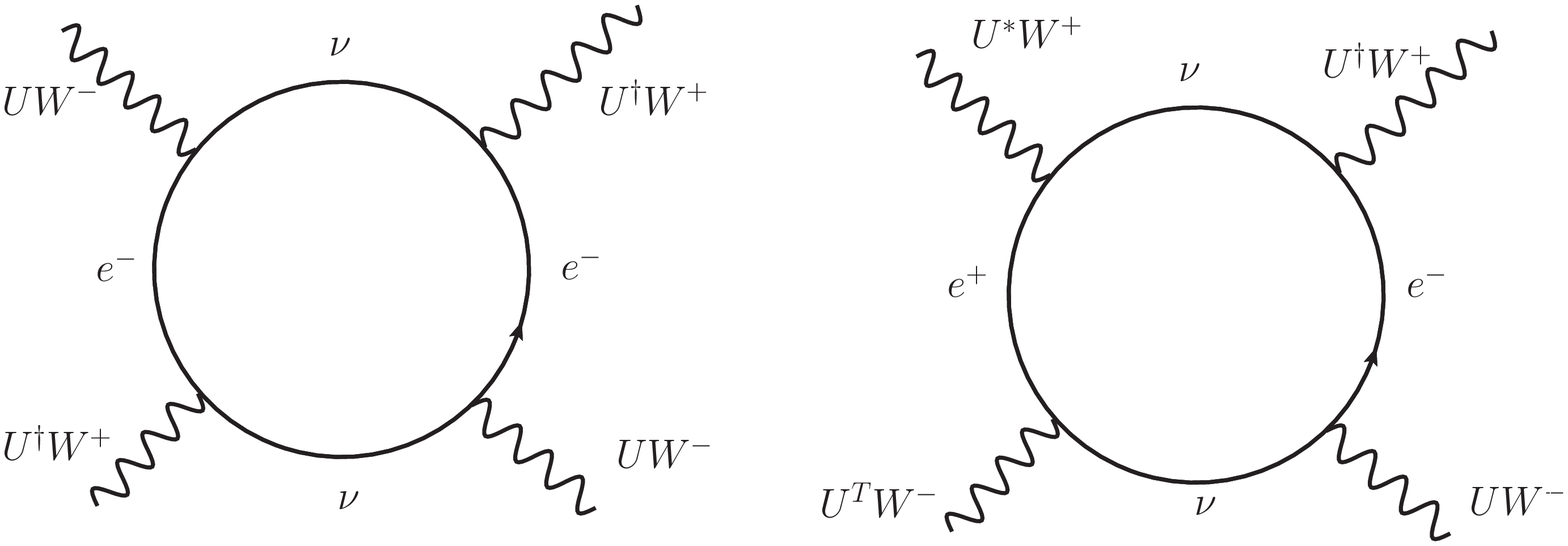,width=120mm}
\caption{Mechanisms involved in $\Gamma_{4a}$ (left panel) and $\Gamma_{4b}$
  (right panel). Gauge bosons are inwards.}
\label{fig:2}
\end{center}
\end{figure}
As said, the first term with a CP odd component is $\Gamma_4$. The two
mechanisms involved there are displayed in Fig.~\ref{fig:2} and they
correspond to two types of momentum integrals, $I_a$ and $I_b$,
\begin{equation}
\begin{split}
I^k_{a,n_e,n_\nu,n_e^\prime,n_\nu^\prime}
&=
\Im \int \frac{d^4 p}{(2\pi)^4} (p^2)^{k/2}
\,
\tr\left(
 N_e^{n_e} N_\nu^{n_\nu}
 N_e^{n_e^\prime} N_\nu^{n_\nu^\prime}
\right)
,
\\
I^k_{b,n_e,n_\nu,n_e^\prime,n_\nu^\prime}
&=
\Im \int \frac{d^4 p}{(2\pi)^4} (p^2)^{k/2}
\,
\tr\left(
 N_e^{n_e} m_\nu N_\nu^{n_\nu}
 N_e^*{}^{n_e^\prime} m_\nu N_\nu^{n_\nu^\prime}
\right)
,
\end{split}
\label{eq:4.8}
\end{equation}
where the exponents $n_e$, $n_\nu$, $n_e^\prime$, $n_\nu^\prime$ are natural
numbers,
\begin{equation}
N_e = U^\dagger \frac{1}{p^2 + m_e^2} U
,\qquad
N_\nu = \frac{1}{p^2 + m_\nu^2}
,
\end{equation}
and $m_e$ and $m_\nu$ denote the positive and diagonal mass matrices of
charged and neutral leptons, respectively. The CP odd sector only makes use of
the imaginary parts of the integrals.  For these integrals, the following
symmetries are easily established
\begin{equation}
\begin{split}
I^k_{a,n_e,n_\nu,n_e^\prime,n_\nu^\prime}
&=
- I^k_{a,n_e^\prime,n_\nu,n_e,n_\nu^\prime}
= - I^k_{a,n_e,n_\nu^\prime,n_e^\prime,n_\nu}
,
\\
I^k_{b,n_e,n_\nu,n_e^\prime,n_\nu^\prime}
&=
+ I^k_{b,n_e^\prime,n_\nu,n_e,n_\nu^\prime}
= - I^k_{b,n_e,n_\nu^\prime,n_e^\prime,n_\nu}
.
\end{split}
\label{eq:4.10a}
\end{equation}

Using the relations (11.6) and (11.7) of \cite{GarciaRecio:2009zp}, the basic
momentum integrals required to obtain $I_a$ and $I_b$ can be reduced to
contour integrals which are easily computed by residues.

\begin{figure}[h]
\begin{center}
\includegraphics[width=0.5\textwidth]{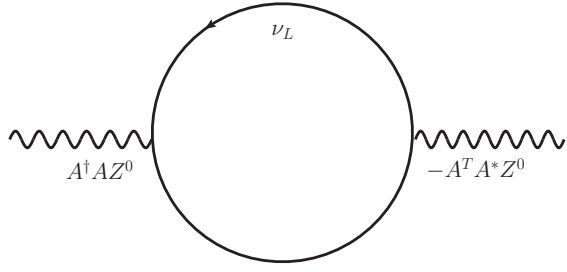}
\caption{Graph contributing to CP violation with only neutral particles,
  provided the neutrinos have mixed Dirac-Majorana masses.}
\label{fig:4}
\end{center}
\end{figure}
Before closing this section, we remark that the necessity of charged gauge
bosons in the fermion loop to produce CP violation follows from the fact that
the complex mass matrices are summarized into $m_e$, $m_\nu$ and $U$. The
first two are real and the latter appears though the operator $\bar{e}U \sW^-
\nu$ and its hermitian conjugate. This is true for pure Dirac or pure
Majorana neutrinos, but it no longer holds in the general case of mixed Dirac
and Majorana masses. Indeed, using the couplings in \eq{3.8} one can construct
CP odd graphs involving no charged boson. One such graph is displayed in
Fig. \ref{fig:4}. The driving operator there is of the type
\begin{equation}
i \epsilon_{\mu\nu\alpha\beta} Z_{\mu\nu} Z_{\alpha\beta} \, \Im \tr ( A^\dagger A
f_1(m_\nu) A^T A^* f_2(m_\nu))
,
\label{eq:4.11}
\end{equation}
where $f_1(m_\nu)$ and $f_2(m_\nu)$ are real and diagonal mass matrices.  This
operator is P odd and of dimension 4, so the coupling is logarithmically UV
divergent. This term can give a CP violating contribution to the $Z^0$
selfenergy for $g\ge 2$ and $N_s\ge 1$. Here $m_\nu$ refers to $m_1$ of
\Eq{3.7}. Similar terms appear with $m_2$ and $B$ at this fourth order. Of
course, in the limiting Dirac or Majorana cases $A$ and $B$ are unitary or
zero and these couplings vanish.

\section{\textsf{Operator K for Majorana neutrinos}}
\label{sec:5}

The case of Dirac fermions has been addressed in \cite{GarciaRecio:2009zp} for
quarks. The results obtained there translate almost immediately to the case of
Dirac neutrinos, therefore we give more details of the derivation of the
Majorana case, which is also more involved.

We consider pure Majorana left-handed neutrinos, that is, $M_D=0$.
Hypothetical right-handed neutrinos can be disregarded since they decouple from
the other fields. In this case, the Lagrangian in the form given in \Eq{2.26},
with the identifications in Eqs. (\ref{eq:3.1}-\ref{eq:3.3}), depends on the
following Dirac operator
\begin{equation}
\begin{split}
\D_M & =
~
\begin{matrix}
\bar{\nu}_L \\ \bar{e}_L \\ \bar{e}^c_R \\ \bar{e}_R \\ \bar{\nu}^c_L \\ \bar{e}_L^c
\end{matrix}
\begin{pmatrix}
0 & M_L^* & 0 & D_\nu + Z & W^+ & 0 \\
M_e & 0 & 0 & W^- & D_e - Z & 0 \\
0 & 0 & M_e^T & 0 & 0 & D_e^* \\
D_e & 0 & 0 & 0 & M_e^\dagger & 0 \\
0 & D_\nu - Z & -W^- & M_L & 0 & 0 \\
0 & -W^+ & D_e^* + Z & 0 & 0 & M_e^* 
\end{pmatrix}
\\
&~~~~~~~
\begin{matrix}
~~~~~ e_R & ~~~~ \nu_L^c & ~~~~~~~ e^c_L &  ~~~~~~ \nu_L & ~~~~~~~ e_L &  ~~~~ e_R^c
\end{matrix}
\end{split}
\label{eq:4.1}
\end{equation}
The fields associated to each row and column of the matrix are also displayed.
The rows and columns associated to the fields $\nu_R$, $\bar\nu_R$, $\nu_R^c$
and $\bar\nu_R^c$ haven been dropped.

$M_L$ and $M_e$ are complex $3\times 3$ matrices which we assume to be regular
(any massless case should be obtained as a limit) and $M_L$ is symmetric. The
diagonalization of these mass matrices is addressed below.

In \Eq{4.1} the contraction of fourvectors with the Dirac gamma
matrices is understood and not explicitly displayed. In addition, to
avoid clumsiness \emph{we have included the Higgs field factors in the mass
  matrices}, so $M_L$ really stands for $(\phi/v)^2M_L$ and $M_e$
stands for $(\phi/v) M_e$. This means that $M_{L,e}$ are not constant,
rather
\begin{equation}
[\partial_\mu, M_L] = 2\varphi_\mu(x) M_L
,\qquad
[\partial_\mu, M_e] = \varphi_\mu(x) M_e
,
\end{equation}
where we have introduced the auxiliary Higgs field
\begin{equation}
\varphi_\mu(x) :=  \frac{\partial_\mu \phi(x)}{\phi(x)}
.
\end{equation}

Before proceeding let us make a small digression. It is clear that $\D_M$ above
contains some redundant information, since the duplication of the charged
lepton field (first $e_{L,R}$ and $\bar{e}_{L,R}$ and then $e_{L,R}^c$ and
$\bar{e}_{L,R}^c$), being of Dirac type, is not strictly necessary. Indeed,
the Lagrangian can be written in matrix form as
\begin{equation}
\begin{split}
\mathcal{L}(x) &=
\begin{pmatrix}
\bar{\nu}_L & \bar{\nu}_L^c & \bar{e}_L &  \bar{e}_R
\end{pmatrix} 
\begin{pmatrix}
\frac{1}{2}(D_\nu + Z) & \frac{1}{2}M_L^* & W^+ & 0 \\
\frac{1}{2}M_L & \frac{1}{2}(D_\nu - Z) & 0 & 0 \\
W^- & 0 & D_e-Z & M_e \\
0 & 0 & M_e^\dagger & D_e 
\end{pmatrix}
\begin{pmatrix}
\nu_L \\ \nu_L^c \\ e_L \\ e_R
\end{pmatrix} 
\\
&= 
\begin{pmatrix}
\bar{\nu}_L & \bar{\nu}_L^c & \bar{e}_L &  \bar{e}_R
\end{pmatrix} 
\begin{pmatrix}
\frac{1}{2}(D_\nu + Z) & \frac{1}{2}m_\nu & U^\dagger W^+ & 0 \\
\frac{1}{2}m_\nu & \frac{1}{2}(D_\nu - Z) & 0 & 0 \\
U W^- & 0 & D_e-Z & m_e \\
0 & 0 & m_e^\dagger & D_e 
\end{pmatrix}
\begin{pmatrix}
\nu_L \\ \nu_L^c \\ e_L \\ e_R
\end{pmatrix}
. 
\end{split}
\label{eq:4.4}
\end{equation}
In the second form $m_e$ and $m_\nu$ are diagonal and positive, namely, by
taking $M_L = U^* m_\nu U^\dagger$ in the basis in which $M_e$ is diagonal.
This form of the Lagrangian is of course correct, however, it involves both
real and complex Grassman fields. As a consequence the effective action
\emph{is not obtained} as the determinant of the associated Dirac operator.
For instance, such determinant does not contain contributions with $U$ and
$U^T$ simultaneously, which are actually present in the effective action
(namely, through the integrals $I_b$ in (\ref{eq:4.8})). These terms are
correctly generated by the Lagrangian in \Eq{4.4} taking into account that the
Wick contraction of $\nu_L$ and $\nu_L^c$ is not vanishing (see
Fig.~\ref{fig:1}):
\begin{figure}[h]
\begin{center}
\includegraphics[width=0.5\textwidth]{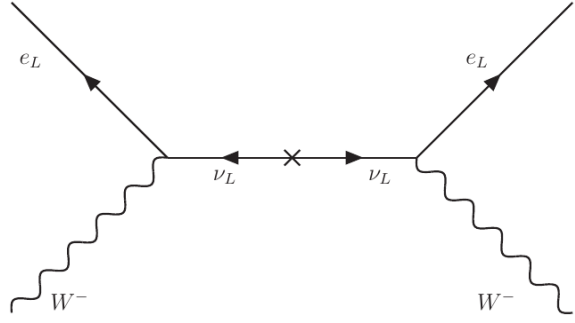}
\caption{Feynman graph corresponding to \Eq{4.5}.}
\label{fig:1}
\end{center}
\end{figure}
\begin{equation}
\hbox{$
( \bar{e}_L U W^- \nu_L) ( \bar{\nu}_L m_\nu \nu_L^c) (\bar{e}_L U W^- \nu_L )
\hspace{-43mm}\underwick{1}{<1{} ~~~~~~~~~~~~~  >1{}}
\hspace{-10mm}\underwick{2}{<1{} ~~~~~~~~~~~~~~~~~~~~~~~~ >1{}}
$}
~~~~~ \sim ~~ \bar{e}_L U  W^- f(m_\nu) U^T  W^- e_L^c
.
\label{eq:4.5}
\end{equation}
On the contrary, within the formalism based on (\ref{eq:4.1}) only
contractions of the type $\psi \bar\psi$ (rather than $\psi\psi$ or $\bar\psi
\bar\psi$) are required. We find it preferable to duplicate all fields thereby
reducing the calculation of the effective action to that of a determinant.

Coming back to (\ref{eq:4.1}), since $\D_M$ is of the general form in \Eq{2.8}
we can proceed to construct the auxiliary operators $K_{L,R}$ of \Eq{2.19}. As
already noted, the two operators contain the same information for UV
convergent contributions to the effective action. In what follows we use
$K_L$. The inverse of $\mrl$ is readily obtained:
\begin{equation}
\mrl^{-1} = \begin{pmatrix}
0 & M_L^{-1} & 0 \\ M_e^\dagger{}^{-1} & 0 & 0 \\ 0 & 0 & M_e^*{}^{-1}
\end{pmatrix}
.
\end{equation}
A straightforward calculation then gives
\small
\begin{equation}
K_L = 
\begin{pmatrix}
M_L^\dagger M_L - (D_\nu +Z )(D_\nu - Z + 2 \varphi) &
-W^+ (D_e + \varphi) &
 (D_\nu + Z) W^- M_L^{-1} M_e^* \\
-W^- (D_\nu - Z + 2\varphi) &
M_e M_e^\dagger -(D_e-Z)(D_e + \varphi) &
W^- W^- M_L^{-1} M_e^* \\
D_e^* W^+ M_e^*{}^{-1} M_L &
0 &
M_e^T M_e^* - D_e^* (D_e^* + Z + \varphi)
\end{pmatrix}
.
\label{eq:4.7}
\end{equation}
\normalsize
In terms of $K_L$, the effective action is given
by
\begin{equation}
\Gamma = - \frac{1}{2} \Tr (P_R \log K_L)
\label{eq:5.8}
\end{equation}
where the factor $1/2$ is that in \Eq{2.27}, and it takes into account that
all fields have been duplicated in order to achieve a Lagrangian formally of
Dirac type (that is, with effective action directly related to the determinant
of the differential operator).

We remark that the matrix elements displayed in (\ref{eq:4.7}) are themselves
$4\times4$ matrices in Dirac space and $g\times g$ matrices in generation
space, for $g$ generations. Also, the fields naturally involved there are
$(\nu_L^c, e_R, e_L^c)$ and $(\bar{\nu}_L , \bar{e}_L ,\bar{e}^c_R ) $. More
precisely, from \Eq{2.24} (adapted to include the duplication required by
Majorana terms) it follows that the propagator of these fields is correctly
reproduced by the effective Lagrangian
\begin{equation}
\mathcal{L}^\text{eff}(x) = 
\begin{pmatrix}
\bar{\nu}_L & \bar{e}_L & \bar{e}_R^c 
\end{pmatrix}
K_L
\begin{pmatrix}
M_L^{-1} \nu_L^c \\ M_e^\dagger{}^{-1} e_R \\ M_e^*{}^{-1} e_L^c 
\end{pmatrix}
.
\end{equation}

Since for the effective action all we needed is $\Det (P_R K_L)$, there
remains the freedom to apply similarity transformations to $K_L$ to obtain a
more convenient form. To this end, we multiply $K_L$ on the left by the matrix
$\phi^{-1}\text{\rm diag}(M_L,M_L,M_e^*)$, and its inverse on the right. This
produces the equivalent operator
\begin{equation}
K_1 = 
\begin{pmatrix}
M_L M_L^\dagger - D_{\nu+} D_{\nu-} &- W^+ D_e & D_{\nu+} W^- \\
-W^- D_{\nu-} & 
M_L M_e M_e^\dagger M_L^{-1} -D_{e-} D_e &
W^- W^- \\
D_e^* W^+ &
0 &
M_e^* M_e^T - D_e^* D_{e+}^*
\end{pmatrix}
,
\label{eq:4.10}
\end{equation}
where we have defined the following shorthands
\begin{equation}
D_{\nu\pm} \equiv D_\nu \pm (Z - \varphi),
\quad
D_{e\pm} \equiv D_e \pm (Z + \varphi),
\quad
D^*_{e\pm} \equiv D_e^* \pm (Z + \varphi)
.
\end{equation}
Below we will also make use of the notation
\begin{equation}
X^{(n)} \equiv \phi^{-n} X \phi^n
,\quad
D^{(n)}_\mu = \phi^{-n} D_\mu \phi^n = D_\mu + n \varphi_\mu
,
\end{equation}
where $X$ is a generic quantity and in particular $D_\mu= \partial_\mu +
V_\mu$ is any derivative operator.

It is noteworthy that in the neutrino sector only the combination $Z-\varphi$
appears, and only $Z+\varphi$ in the charged sector, and moreover, the same
property holds true for the Dirac case \cite{GarciaRecio:2009zp}. We
have no a priori explanation for this regularity.
One should note however that this property needs not translate immediately
to the effective action since new $\varphi_\mu$ can be generated through the
$\phi$ dependence contained in $M_L$ and $M_e$.

For comparison, we quote here the similar matrix $K_L$ for Dirac neutrinos
\cite{GarciaRecio:2009zp}\footnote{We have applied a similarity transformation
  $K \to \phi K \phi^{-1}$ to the matrix of Eq. (7.1) of
  \cite{GarciaRecio:2009zp}, in order to enforce the dependence $Z\pm\varphi$
  for the neutrino (up) and charged (down) sectors.}:
\begin{equation}
K_\text{\rm Dirac} = 
\begin{pmatrix}
M_D M_D^\dagger - D_{\nu+} D_{\nu} &- W^+ D_e \\
-W^- D_{\nu} & 
M_e M_e^\dagger - D_{e-} D_e \\
\end{pmatrix}
,
\label{eq:5.13}
\end{equation}
where $D_\nu=\partial$ and the Dirac neutrino mass matrix $M_D$ includes a
factor $\phi/v$. When this is regarded as an effective Lagrangian, the fermion
fields involved in this matrix are $(\bar{\nu}_L,\bar{e}_L)$ and
$\displaystyle \begin{pmatrix} M_D^{-1}\nu_R \\ M_e^{-1}e_R\end{pmatrix}$.

To proceed with the Majorana case we bring the fermion masses to a
diagonal form:
\begin{equation}
M_L = A m_\nu A^T ,\quad
M_e = B m_e C
,\quad
U:= B^\dagger A^*
.
\label{eq:5.14}
\end{equation}
Here $A$, $B$, $C$ are suitable constant unitary matrices such that $m_\nu$
and $m_e$ are positive diagonal matrices (which include factors $(\phi/v)^2$
and $\phi/v$, respectively). Multiplying $K_1$ in (\ref{eq:4.10}) by
$\text{\rm diag} (A^\dagger, A^\dagger , A^\dagger)$ on the left, and its
inverse on the right, produces the equivalent matrix (also denoted by $K_1$)
\begin{equation}
K_1 = 
\begin{pmatrix}
m^2_\nu - D_{\nu+} D_{\nu-} & - W^ + D_e & D_{\nu+} W^- \\
-W^- D_{\nu-} & 
m_\nu U^\dagger m_e^2 U m_\nu^{-1} - D_{e-} D_e &
W^- W^- \\
D_e^* W^+ &
0 &
U^T m_e^2 U^* - D_e^* D_{e+}^*
\end{pmatrix}
.
\label{eq:4.15}
\end{equation}
The determinant of $K_1$ is unchanged by transformations of $U$ of the type
\begin{equation}
U \to \text{\rm diag}(e^{\theta_1},\cdots,e^{\theta_g}) U 
\text{\rm diag}(\pm,\ldots,\pm)
,
\end{equation}
and these are the unique allowed transformations if masses are not
degenerated.

If the expression in (\ref{eq:4.15}) is used directly in the calculation of
the effective action, inverse powers of $m_\nu$ appear in intermediate
steps, although eventually they can be removed in every single case. In order
to obtain directly the expressions without $m_\nu^{-1}$, we proceed as
follows. First we define the following propagators
\begin{equation}
\begin{split}
G_\nu &:= \left( m_\nu^2- D_{\nu+} D_{\nu-} \right)^{-1} ,\\
G_e &:= U^\dagger \left( m_e^2 - D_{e-} D_e \right)^{-1} U ,\\
G_e^* &:= U^T \left(m_e^2 - D_e^* D_{e+}^* \right)^{-1} U^*
.
\end{split}
\end{equation}
In terms of these, the matrix $K_1$ can be expressed as
\begin{equation}
K_1 = 
\begin{pmatrix}
G_\nu^{-1} & - W^ + D_e & D_{\nu+} W^- \\
-W^- D_{\nu-} & 
m_\nu G_e^{(2)}{}^{-1} m_\nu^{-1} &
W^- W^- \\
D_e^* W^+ &
0 &
G_e^*{}^{-1}
\end{pmatrix}
.
\label{eq:4.18}
\end{equation}
Here $G_e^{(2)} = \phi^{-2} G_e \phi^2$. This amounts to shifting $D_e \to
D_e^{(2)} = D_e + 2\varphi$, and comes about from $D_{e-}D_e = m_\nu
D_{e-}^{(2)}D_e^{(2)} m_\nu^{-1}$ in (\ref{eq:4.15}). Next, we define a new
matrix $K_2$ by appending to $K_1$ a factor on the right
\begin{equation}
K_2 = K_1  
\begin{pmatrix}
1 & 0 & - G_\nu D_{\nu+} W^- \\
0 & 1 & 0 \\
0 & 0 & 1 
\end{pmatrix}
.
\end{equation}
Clearly, the determinant of the appended factor is unity, hence 
$\Det\,K_2 = \Det \,K_1$, and the effective action remains unchanged. An
explicit calculation produces:
\begin{equation}
K_2 = 
\begin{pmatrix}
G_\nu^{-1} & - W^ + D_e & 0 \\
-W^- D_{\nu-} & m_\nu G_e^{(2)}{}^{-1} m_\nu^{-1} &
m_\nu W^- G^\prime_\nu W^- m_\nu \\
D_e^* W^+ & 0 & G_e^*{}^{-1} - D_e^* W^+ G_\nu D_{\nu+} W^-
\end{pmatrix}
,
\end{equation}
where we have introduced a modified neutrino propagator
\begin{equation}
G^\prime_\nu := 
\left( m_\nu^2 - D_{\nu-}^{(-2)}D_{\nu+}^{(2)} \right)^{-1}
,
\end{equation}
and we have made use of the identity
\begin{equation}
1 + D_{\nu-} G_\nu D_{\nu+} = m_\nu G^\prime_\nu m_\nu
.
\end{equation}

The sought for form of $K$ with no inverse powers of $m_\nu$ is obtained by
multiplying $K_2$ by $\text{\rm diag}(m_\nu^{-1},m_\nu^{-1},1)$ on the left,
and its inverse on the right:
\begin{equation}
K := 
\begin{pmatrix}
G_\nu^{(2)}{}^{-1} & - W^+ D_e^{(2)} & 0 \\
-W^- D_{\nu-}^{(2)} & G_e^{(2)}{}^{-1} &
W^- G^\prime_\nu W^- m_\nu \\
D_e^* W^+ m_\nu & 0 & G_e^*{}^{-1} - D_e^* W^+ G_\nu D_{\nu+} W^-
\end{pmatrix}
.
\label{eq:4.23}
\end{equation}
This is our final form of the operator $K$ for Majorana neutrinos.  Let us
emphasize that $K$ provides the chiral invariant part of the lepton-induced
effective action for any number of $W$'s and for all sectors, P even or odd
and CP even or odd, and its use is not restricted to a derivative expansion.

We note that all manipulations used above contain only blocks with an even
number of Dirac matrices, so no problem arises from the presence of the factor
$P_R$ in the trace in \neq{5.8}. The same remark applies for the next section.

\section{\textsf{Effective action in the CP odd sector}}
\label{sec:6}

\subsection{\textsf{$\Gamma_4$ for Majorana and Dirac neutrinos}}

The effective action is given by
\begin{equation}
\Gamma = - \frac{1}{2} \Tr (P_R \log K)
.
\end{equation}
In order to use this form for the CP odd sector we will expand the right-hand
side in powers of $W^\pm$. To this end let us express $K$ in the form
\begin{equation}
K = K_0 ( 1 - \Delta_1 - \Delta_2 )
,
\end{equation}
with
\begin{equation}
\begin{split}
K_0 &= \begin{pmatrix}
G_\nu^{(2)}{}^{-1} &  & 0 \\
0 & G_e^{(2)}{}^{-1} & 0  \\
0 & 0 & G_e^*{}^{-1}
\end{pmatrix}
,
\\
\Delta_1 &= \begin{pmatrix}
0 & G_\nu^{(2)} W^+ D_e^{(2)} & 0 \\
G_e^{(2)} W^- D_{\nu-}^{(2)} & 0 & 0  \\
- G_e^* D_e^* W^+ m_\nu & 0 & 0
\end{pmatrix}
,\quad
\Delta_2 = \begin{pmatrix}
0 & 0 & 0 \\
0 & 0 & - G_e^{(2)} W^- G^\prime_\nu W^- m_\nu \\
0 & 0 & G_e^* D_e^* W^+ G_\nu D_{\nu+} W^-
\end{pmatrix}
.
\end{split}\end{equation}
In this way (once again using formal manipulations which are justified at the
order we are working, due to UV convergence)
\begin{equation}
\Tr(P_R \log K) = \Tr(P_R \log K_0) - \sum_{n=1}^\infty \frac{1}{n}
\Tr (P_R (  \Delta_1 + \Delta_2)^n )
.
\end{equation}

As has been shown in Sec. \ref{sec:4}, the first term that can contribute to
CP violation is of order $4$ in powers of $W^\pm$. This selects the terms
(with the notation $[\Delta_1]_{1,2}=G_\nu^{(2)} W^+ D_e^{(2)}$, etc)
\begin{equation}
\begin{split}
 \Tr(P_R \log K)_ 4 &= -  \Tr\left(P_R \left ( \frac{1}{4}\Delta_1^4
+  \frac{1}{2} \Delta_2^2 
+ \Delta_1^2 \Delta_2 
\right) \right)
\\
 &= -  \Tr\left(P_R \left ( \frac{1}{2} ([\Delta_1]_{1,2} [\Delta_1]_{2,1} )^2
+ \frac{1}{2} ( [\Delta_2]_{3,3})^2
+ [\Delta_1]_{1,2} [\Delta_2]_{2,3} [\Delta_1]_{3,1}
 \right) \right)
,
\end{split}
\end{equation}
Explictly,
\begin{equation}
\begin{split}
\Gamma_4 &=
\frac{1}{2} \Tr\Big( P_R \Big[ 
\frac{1}{2} \big( G_e^{(2)} W^- D_{\nu-}^{(2)} G_\nu^{(2)} W^+ D_e^{(2)} \big)^2
+ \frac{1}{2} \big( G_e^* D_e^* W^+ G_\nu D_{\nu+} W^- \big)^2
\\ & ~~
+ G_\nu^{(2)} W^+ D_e^{(2)} G_e^{(2)} W^- G^\prime_\nu m_\nu W^- G_e^* D_e^* W^+ m_\nu 
 \Big] \Big)
.
\end{split}
\end{equation}

The two terms with factors one half have a similar structure and they are
actually equal. This can be shown by applying transposition to the second
term, plus the relations
\begin{equation}
\begin{split}
\sD_e^*{}^{(n)}{}^T &= C^{-1} \sD_e^{(-n)} C, \quad
\sD_{\nu\pm}^{(n)}{}^T = C^{-1} \sD_{\nu\mp}^{(-n)} C
,\quad
\sW^\pm{}^T = - C^{-1} \sW^\pm C
,
\\
G_e^*{}^{(n)}{}^T &= C^{-1} G_e^{(-n)} C
, \quad
G_\nu^{(n)}{}^T = C^{-1} G_\nu^{(-n)} C
.
\end{split}
\end{equation}
In summary, for \emph{Majorana neutrinos}, the terms of the effective action
with exactly four charged gauge bosons can be expressed as
\begin{equation}
\Gamma_{4,M} =
\frac{1}{2} \Tr \, P_R \Big[
\big( G_e \sW^- \sD_{\nu-} G_\nu \sW^+ \sD_e \big)^2
+ 
G_e^{(2)} \sW^- G^\prime_\nu \, m_\nu \sW^- G_e^* \sD_e^* \sW^+ G_\nu \sW^+ \sD_e \, m_\nu
 \Big]
.
\label{eq:6.8}
\end{equation}

The similar expression for \emph{Dirac neutrinos} is instead
\begin{equation}
\begin{split}
\Gamma_{4,D} &=
\frac{1}{2} \Tr \, P_R \Big[
\big( G_e \sW^- \sD_\nu \, G^D_\nu \sW^+ \sD_e \big)^2
 \Big]
.
\end{split}
\end{equation}
where
\begin{equation}
G^D_\nu := \left(m_\nu^2 - \sD_{\nu+} \sD_\nu \right)^{-1}
.
\end{equation}
Here $m_\nu$ stands for $(\phi/v)m_\nu$ and it is obtained from $M_D = A m_\nu D$
($A$ and $D$ being unitary matrices). $G_e$ is the same as before, with $U
= B^\dagger A^*$ and $M_e = B m_e C $, as in \Eq{5.14}.

The effective action for Majorana neutrinos contains two types of terms
\begin{equation}
\Gamma_{4,M} = \Gamma_{4,a} + \Gamma_{4,b}
.
\label{eq:6.8a}
\end{equation}

The terms in $\Gamma_{4,a}$ follow a mechanism similar to that of the Dirac
case, namely, the charged bosons alternate along the fermion loop,
$W^-W^+W^-W^+$ (see Fig. \ref{fig:2}, left panel). In fact, the expressions of
$\Gamma_{4,a}$ and $\Gamma_{4,D}$ are identical for contributions not
involving $Z_\mu$ nor $\varphi_\mu$ and so in this case they give the same
contributions.\footnote{This refers to the explicit $Z_\mu$. The $Z^0$ field
  appears also in $D_e$ together with the photon field,
  \Eq{3.5}. Unfortunately, the Dirac results in \cite{GarciaRecio:2009zp}
  cannot be directly adapted to $\Gamma_{4,a}$ when Higgs or $Z^0$ are present
  by means of some clever redefinition of the fields there.} On the other
hand, the mechanism in $\Gamma_{4,b}$ is of the type $W^-W^-W^+W^+$ (see
Fig. \ref{fig:2}, right panel) thereby violating fermionic number
conservation. This is the mechanism responsible for neutrinoless double beta
decay. The presence of such mechanism is particularly clear in the operator
$K$ from the matrix element $[K]_{2,3}= \sW^- G^\prime_\nu \sW^- m_\nu$ in
(\ref{eq:4.23}).

It is also noteworthy that, from CPT invariance one expects that $W^+$ and
$W^-$ should play similar roles. The symmetry between $W^+$ and $W^-$ is
explicit in the Dirac operator of \Eq{4.1} but it is not manifest in $K$. As
noted we could have started from $K_R$. In this case the roles of the charged
bosons would be reversed. The symmetry is restored in $\Gamma_4$ and certainly
in the final results.

\subsection{\textsf{Method of covariant symbols}}

The actual calculation of the terms $\Gamma_{4+2}$ has been done using the
method of covariant symbols \cite{Pletnev:1998yu,MoralGamez:2011en}. Quite
simply, for any operator of the type $f(D,M)$, where the $D$ are covariant
derivatives and the $M(x)$ are matrices in internal space, such as that in
\Eq{6.8}, the functional trace can be expressed as
\begin{equation}
\Tr \, f(D,M) = \int \frac{d^dx d^dp}{(2\pi)^d} \,\tr f(\bar{D},\bar{M})
.
\label{eq:6.12a}
\end{equation}
$\tr$ refers to internal degrees of freedom, $p_\mu$ is the fermion loop
momentum, and $\bar{D}$ and $\bar{M}$ are the \emph{covariant symbols} of $D$
and $M$. These are gauge covariant operators which are \emph{multiplicative}
with respect to $x$ and contain derivatives with respect to $p$, namely,
\begin{equation}\begin{split}
  \bar{M} &= 
\sum_{n=0}^\infty \frac{i^n}{n!} 
\left( \hat{D}_{\alpha_1} \cdots \hat{D}_{\alpha_n} M \right)
\partial^p_{\alpha_1} \cdots \partial^p_{\alpha_n} ,
\\
\bar{D}_\mu &=
i p_\mu + \sum_{n=1}^\infty \frac{ i^n n }{(n+1)!}
\left( \hat{D}_{\alpha_1} \cdots \hat{D}_{\alpha_n} D_\mu \right)
\partial^p_{\alpha_1} \cdots \partial^p_{\alpha_n}
,
\end{split}
\label{eq:6.12}
\end{equation}
with $\partial^p_\alpha = \partial / \partial p_\alpha$ and $\hat{D}_\alpha X
= [D_\alpha,X]$, in particular, $\hat{D}_\alpha D_\mu = F_{\alpha\mu}$. The
crucial property of the covariant symbols, besides being manifestly gauge
covariant and multiplicative as operators, is that they define a
representation of the algebra of operators, that is, $\overline{f(X,Y)} =
f(\bar{X},\bar{Y})$. Therefore, one can simply compute the symbols of the
basic blocks and use them in the full expression.

In our case an application of the method of covariant symbols in
(\ref{eq:6.8}) produces
\begin{equation}
\begin{split}
\Gamma_{4,a} &=  \int \frac{d^4x d^4p}{(2\pi)^4}
\frac{1}{2} \tr \Big[ P_R \,
\left( \bar G_e \bar\sW^- \bar\sD_{\nu-} \bar G_\nu \bar\sW^+ \bar\sD_e \right)^2
\Big]
\\
\Gamma_{4,b} &=  \int \frac{d^4x d^4p}{(2\pi)^4}
\frac{1}{2} \tr \Big[ P_R \, 
\bar G_e^{(2)} \bar \sW^- \bar G^\prime_\nu \, \bar m_\nu \bar\sW^- \bar G_e^*
\bar\sD_e^* \bar\sW^+ \bar G_\nu \bar\sW^+ \bar\sD_e \, \bar m_\nu
 \Big]
,
\end{split}
\label{eq:6.8s}
\end{equation}
where $\tr$ refers to Dirac and flavor spaces. The covariant symbols of
the basics operators in (\ref{eq:6.8s}) are displayed in Appendix \ref{app:A}.

We want to work out the leading CP violating terms, that is, those driven by
operators of lowest dimension, which are those in $\Gamma_{4+2}$. Therefore,
in (\ref{eq:6.8s}) we select contributions with exactly two derivatives, where
each $\hat{D}_\mu$, $Z_\mu$ or $\varphi_\mu$ counts as one derivative.

Using the covariant symbols in Appendix \ref{app:A}, the calculation of the
effective action proceeds from (\ref{eq:6.8s}) by i) removing all momentum
derivatives, applying them either on the right or the left, ii) taking an
angular average over the momenta, iii) evaluating the Dirac gamma traces, iv)
factoring each term into a momentum integral (involving the mass matrices) and
an operator (involving the external fields and their derivatives), v)
rearranging indices in the operators, including Bianchi identities (namely, $[
  \hat{D}_\alpha,\hat{D}_\beta ] X = [F_{\alpha\beta},X]$), vi) using
integration by parts, and vii) using identities between momentum integrals to
simplify the final result.  At step iv) the CP odd terms can be already
isolated by selecting the antihermitian/hermitian part of the operators in the
P even/odd sectors. The calculation has been repeated using the method of
ordinary symbols \cite{Nepomechie:1984wt,Salcedo:1994qy} as a check of the
results.

\subsection{\textsf{Allowed operators and their couplings}
\label{sec:6.3}
}

\begin{table}[h]
\caption{List of P even and CP odd operators of the type $4+2$. }
\centering
\begin{tabular}{|ll|ll|}
\hline
 $ \OpA_{ 1 } \EQb $ & $ W^+_{ \alpha } W^+_{ \alpha } W^-_{ \beta \beta } W^-_{ \gamma \gamma } - \text{c.c.}  $  & 
 $ \OpA_{ 21 } \EQb $ & $ W^+_{ \alpha } W^+_{ \beta } W^-_{ \alpha } W^-_{ \beta } Z_{ \gamma \gamma } $  \\
 $ \OpA_{ 2 } \EQb $ & $ W^+_{ \alpha } W^+_{ \alpha } W^-_{ \beta \gamma } W^-_{ \beta \gamma } - \text{c.c.}  $  & 
 $ \OpA_{ 22 } \EQb $ & $ W^+_{ \alpha } W^+_{ \beta } W^-_{ \alpha } W^-_{ \gamma } Z_{ \beta \gamma } - \text{c.c.}  $  \\
 $ \OpA_{ 3 } \EQb $ & $ W^+_{ \alpha } W^+_{ \alpha } W^-_{ \beta \gamma } W^-_{ \gamma \beta } - \text{c.c.}  $  & 
 $ \OpA_{ 23 } \EQb $ & $ W^+_{ \alpha \beta } W^+_{ \gamma } W^-_{ \alpha } W^-_{ \beta } Z_{ \gamma } - \text{c.c.}  $  \\
 $ \OpA_{ 4 } \EQb $ & $ W^+_{ \alpha } W^+_{ \beta } W^-_{ \alpha \beta } W^-_{ \gamma \gamma } - \text{c.c.}  $  & 
 $ \OpA_{ 24 } \EQb $ & $ W^+_{ \alpha } W^+_{ \beta } W^-_{ \alpha \beta } W^-_{ \gamma } \varphi_{ \gamma } - \text{c.c.}  $  \\
 $ \OpA_{ 5 } \EQb $ & $ W^+_{ \alpha } W^+_{ \beta } W^-_{ \alpha \gamma } W^-_{ \beta \gamma } - \text{c.c.}  $  & 
 $ \OpA_{ 25 } \EQb $ & $ W^+_{ \alpha \beta } W^+_{ \gamma } W^-_{ \alpha } W^-_{ \gamma } Z_{ \beta } - \text{c.c.}  $  \\
 $ \OpA_{ 6 } \EQb $ & $ W^+_{ \alpha } W^+_{ \beta } W^-_{ \alpha \gamma } W^-_{ \gamma \beta } - \text{c.c.}  $  & 
 $ \OpA_{ 26 } \EQb $ & $ W^+_{ \alpha } W^+_{ \beta } W^-_{ \alpha \gamma } W^-_{ \beta } \varphi_{ \gamma } - \text{c.c.}  $  \\
 $ \OpA_{ 7 } \EQb $ & $ W^+_{ \alpha } W^+_{ \beta } W^-_{ \gamma \alpha } W^-_{ \gamma \beta } - \text{c.c.}  $  & 
 $ \OpA_{ 27 } \EQb $ & $ W^+_{ \alpha \beta } W^+_{ \beta } W^-_{ \alpha } W^-_{ \gamma } Z_{ \gamma } - \text{c.c.}  $  \\
 $ \OpA_{ 8 } \EQb $ & $ W^+_{ \alpha \alpha } W^+_{ \beta } W^-_{ \beta \gamma } W^-_{ \gamma } - \text{c.c.}  $  & 
 $ \OpA_{ 28 } \EQb $ & $ W^+_{ \alpha } W^+_{ \beta } W^-_{ \alpha \gamma } W^-_{ \gamma } \varphi_{ \beta } - \text{c.c.}  $  \\
 $ \OpA_{ 9 } \EQb $ & $ W^+_{ \alpha \alpha } W^+_{ \beta } W^-_{ \gamma \beta } W^-_{ \gamma } - \text{c.c.}  $  & 
 $ \OpA_{ 29 } \EQb $ & $ W^+_{ \alpha \beta } W^+_{ \gamma } W^-_{ \beta } W^-_{ \gamma } Z_{ \alpha } - \text{c.c.}  $  \\
 $ \OpA_{ 10 } \EQb $ & $ W^+_{ \alpha \beta } W^+_{ \alpha } W^-_{ \beta \gamma } W^-_{ \gamma } - \text{c.c.}  $  & 
 $ \OpA_{ 30 } \EQb $ & $ W^+_{ \alpha } W^+_{ \beta } W^-_{ \gamma \alpha } W^-_{ \beta } \varphi_{ \gamma } - \text{c.c.}  $  \\
 $ \OpA_{ 11 } \EQb $ & $ W^+_{ \alpha \beta } W^+_{ \gamma } W^-_{ \beta \gamma } W^-_{ \alpha } - \text{c.c.}  $  & 
 $ \OpA_{ 31 } \EQb $ & $ W^+_{ \alpha \beta } W^+_{ \alpha } W^-_{ \beta } W^-_{ \gamma } Z_{ \gamma } - \text{c.c.}  $  \\
 $ \OpA_{ 12 } \EQb $ & $ W^+_{ \alpha } W^+_{ \alpha } W^-_{ \beta } W^-_{ \beta } Z_{ \gamma \gamma } $  & 
 $ \OpA_{ 32 } \EQb $ & $ W^+_{ \alpha } W^+_{ \beta } W^-_{ \gamma \alpha } W^-_{ \gamma } \varphi_{ \beta } - \text{c.c.}  $  \\
 $ \OpA_{ 13 } \EQb $ & $ W^+_{ \alpha } W^+_{ \alpha } W^-_{ \beta } W^-_{ \gamma } Z_{ \beta \gamma } - \text{c.c.}  $  & 
 $ \OpA_{ 33 } \EQb $ & $ W^+_{ \alpha \alpha } W^+_{ \beta } W^-_{ \beta } W^-_{ \gamma } Z_{ \gamma } - \text{c.c.}  $  \\
 $ \OpA_{ 14 } \EQb $ & $ W^+_{ \alpha } W^+_{ \alpha } W^-_{ \beta } W^-_{ \gamma } \varphi_{ \beta \gamma } - \text{c.c.}  $  & 
 $ \OpA_{ 34 } \EQb $ & $ W^+_{ \alpha } W^+_{ \beta } W^-_{ \gamma \gamma } W^-_{ \alpha } \varphi_{ \beta } - \text{c.c.}  $  \\
 $ \OpA_{ 15 } \EQb $ & $ W^+_{ \alpha \alpha } W^+_{ \beta } W^-_{ \gamma } W^-_{ \gamma } Z_{ \beta } - \text{c.c.}  $  & 
 $ \OpA_{ 35 } \EQb $ & $ W^+_{ \alpha } W^+_{ \alpha } W^-_{ \beta } W^-_{ \beta } Z_{ \gamma } \varphi_{ \gamma } $  \\
 $ \OpA_{ 16 } \EQb $ & $ W^+_{ \alpha } W^+_{ \alpha } W^-_{ \beta \beta } W^-_{ \gamma } \varphi_{ \gamma } - \text{c.c.}  $  & 
 $ \OpA_{ 36 } \EQb $ & $ W^+_{ \alpha } W^+_{ \alpha } W^-_{ \beta } W^-_{ \gamma } Z_{ \beta } Z_{ \gamma } - \text{c.c.}  $  \\
 $ \OpA_{ 17 } \EQb $ & $ W^+_{ \alpha \beta } W^+_{ \alpha } W^-_{ \gamma } W^-_{ \gamma } Z_{ \beta } - \text{c.c.}  $  & 
 $ \OpA_{ 37 } \EQb $ & $ W^+_{ \alpha } W^+_{ \alpha } W^-_{ \beta } W^-_{ \gamma } Z_{ \beta } \varphi_{ \gamma } - \text{c.c.}  $  \\
 $ \OpA_{ 18 } \EQb $ & $ W^+_{ \alpha } W^+_{ \alpha } W^-_{ \beta \gamma } W^-_{ \beta } \varphi_{ \gamma } - \text{c.c.}  $  & 
 $ \OpA_{ 38 } \EQb $ & $ W^+_{ \alpha } W^+_{ \alpha } W^-_{ \beta } W^-_{ \gamma } \varphi_{ \beta } \varphi_{ \gamma } - \text{c.c.}  $  \\
 $ \OpA_{ 19 } \EQb $ & $ W^+_{ \alpha \beta } W^+_{ \beta } W^-_{ \gamma } W^-_{ \gamma } Z_{ \alpha } - \text{c.c.}  $  & 
 $ \OpA_{ 39 } \EQb $ & $ W^+_{ \alpha } W^+_{ \beta } W^-_{ \alpha } W^-_{ \beta } Z_{ \gamma } \varphi_{ \gamma } $  \\
 $ \OpA_{ 20 } \EQb $ & $ W^+_{ \alpha } W^+_{ \alpha } W^-_{ \beta \gamma } W^-_{ \gamma } \varphi_{ \beta } - \text{c.c.}  $  & 
 $ \OpA_{ 40 } \EQb $ & $ W^+_{ \alpha } W^+_{ \beta } W^-_{ \alpha } W^-_{ \gamma } Z_{ \beta } \varphi_{ \gamma } - \text{c.c.}  $  \\
\hline
\end{tabular}\\

\label{tab:2}
\end{table}
\begin{table}[h]
\caption{List of P odd and CP odd operators of the type $4+2$. }
\centering
\begin{tabular}{|ll|ll|}
\hline
 $ \OpH_{ 1 } \EQb $  &  $ W^+_{ a } W^+_{ a s } W^-_{ a } W^-_{ a s } $  & 
 $ \OpH_{ 14 } \EQb $  &  $ W^+_{ a } W^+_{ s } W^-_{ a } W^-_{ s } Z_{ a a } $  \\ 
 $ \OpH_{ 2 } \EQb $  &  $ W^+_{ a } W^+_{ a s } W^-_{ a } W^-_{ s a } + \text{c.c.}  $  & 
 $ \OpH_{ 15 } \EQb $  &  $ W^+_{ a } W^+_{ a a } W^-_{ a } W^-_{ s } Z_{ s } + \text{c.c.}  $  \\ 
 $ \OpH_{ 3 } \EQb $  &  $ W^+_{ a } W^+_{ s a } W^-_{ a } W^-_{ s a } $  & 
 $ \OpH_{ 16 } \EQb $  &  $ W^+_{ a } W^+_{ s } W^-_{ a } W^-_{ a a } \varphi_{ s } + \text{c.c.}  $  \\ 
 $ \OpH_{ 4 } \EQb $  &  $ W^+_{ a } W^+_{ a a } W^-_{ a } W^-_{ s s } + \text{c.c.}  $  & 
 $ \OpH_{ 17 } \EQb $  &  $ W^+_{ a } W^+_{ s } W^-_{ a } W^-_{ a s } Z_{ a } + \text{c.c.}  $  \\ 
 $ \OpH_{ 5 } \EQb $  &  $ W^+_{ a } W^+_{ a s } W^-_{ s } W^-_{ a a } + \text{c.c.}  $  & 
 $ \OpH_{ 18 } \EQb $  &  $ W^+_{ a } W^+_{ a s } W^-_{ a } W^-_{ s } \varphi_{ a } + \text{c.c.}  $  \\ 
 $ \OpH_{ 6 } \EQb $  &  $ W^+_{ a } W^+_{ s a } W^-_{ s } W^-_{ a a } + \text{c.c.}  $  & 
 $ \OpH_{ 19 } \EQb $  &  $ W^+_{ a } W^+_{ s } W^-_{ a } W^-_{ s a } Z_{ a } + \text{c.c.}  $  \\ 
 $ \OpH_{ 7 } \EQb $  &  $ W^+_{ s } W^+_{ a a } W^-_{ s } W^-_{ a a } $  & 
 $ \OpH_{ 20 } \EQb $  &  $ W^+_{ a } W^+_{ s a } W^-_{ a } W^-_{ s } \varphi_{ a } + \text{c.c.}  $  \\ 
 $ \OpH_{ 8 } \EQb $  &  $ W^+_{ a } W^+_{ a a } W^-_{ s } W^-_{ a s } + \text{c.c.}  $  & 
 $ \OpH_{ 21 } \EQb $  &  $ W^+_{ a } W^+_{ a a } W^-_{ s } W^-_{ s } Z_{ a } + \text{c.c.}  $  \\ 
 $ \OpH_{ 9 } \EQb $  &  $ W^+_{ a } W^+_{ a a } W^-_{ s } W^-_{ s a } + \text{c.c.}  $  & 
 $ \OpH_{ 22 } \EQb $  &  $ W^+_{ a } W^+_{ a a } W^-_{ s } W^-_{ s } \varphi_{ a } + \text{c.c.}  $  \\ 
 $ \OpH_{ 10 } \EQb $  &  $ W^+_{ a a } W^+_{ a a } W^-_{ s } W^-_{ s } + \text{c.c.}  $  & 
 $ \OpH_{ 23 } \EQb $  &  $ W^+_{ a } W^+_{ s } W^-_{ s } W^-_{ a a } Z_{ a } + \text{c.c.}  $  \\ 
 $ \OpH_{ 11 } \EQb $  &  $ W^+_{ a } W^+_{ s } W^-_{ a a } W^-_{ a s } + \text{c.c.}  $  & 
 $ \OpH_{ 24 } \EQb $  &  $ W^+_{ a } W^+_{ s } W^-_{ s } W^-_{ a a } \varphi_{ a } + \text{c.c.}  $  \\ 
 $ \OpH_{ 12 } \EQb $  &  $ W^+_{ a } W^+_{ s } W^-_{ a a } W^-_{ s a } + \text{c.c.}  $  & 
 $ \OpH_{ 25 } \EQb $  &  $ W^+_{ a } W^+_{ s } W^-_{ a } W^-_{ s } Z_{ a } \varphi_{ a } $  \\ 
 $ \OpH_{ 13 } \EQb $  &  $ F^e_{ a a } W^+_{ a } W^+_{ s } W^-_{ a } W^-_{ s } $  & & \\ 
\hline
\end{tabular}\\

\label{tab:3}
\end{table}
To express the results we introduce two bases of CP odd operators, one with
parity even operators and another with parity odd ones. They are displayed in
Tables \ref{tab:2} and \ref{tab:3}. In the operators of the parity odd sector,
the labels $a$ and $s$ denote antisymmetric and symmetric Lorentz indices,
respectively. So for instance
\begin{equation}
W^+_a W^+_s W^-_a W^-_s Z_a \, \varphi_a \equiv \epsilon_{\mu\nu\alpha\beta}
W^+_\mu W^+_\rho W^-_\nu W^-_\rho Z_\alpha \, \varphi_\beta
.
\end{equation}
Operators with $W^\pm$ carrying more than one derivative have been excluded
from the bases, as those operators can be eliminated through integration by
parts.

Identities from integration by parts exist among the operators.  To establish
such relations, one should take into account that the Higgs field $\phi$ is
present in the momentum integrals, through the mass terms, and this may
produce new $\varphi_\mu$ dependences not explicit in \Eq{6.8},
\begin{equation}
0 = \int d^4 x \,\tr \hat{D}_\mu (I \mathcal{O}_\mu )
= \int d^4 x \, \tr \left( \varphi_\mu \phi \frac{\partial I}{\partial\phi} 
\mathcal{O}_\mu
 +  I \hat{D}_\mu\mathcal{O}_\mu \right)
.
\label{eq:6.16}
\end{equation}
To make things simpler, we have assumed that the coupling between mass and
Higgs is predominantly of the type $m_i \to \phi m_i$ (Dirac particles). In
this case $\phi \, (\partial I / \partial\phi) = -2 I$ for operators of
dimension $6$. It should not go unnoticed that this is an approximation taken
on otherwise exact relations, imposed on us by the need to avoid cumbersome
expressions. It only affects the Majorana neutrino case in terms with
$\varphi_\mu$.

With this proviso, the following by-parts integration relations are found among
the P even operators
\begin{equation}
\begin{split}
0 &= \OpA_{1} - \OpA_{3} - 2 \OpA_{8} + 2 \OpA_{10} - 2 \OpA_{16} + 2 \OpA_{18}
,\\
0 &= \OpA_{4} - \OpA_{6} - \OpA_{9} + \OpA_{11} + 2 \OpA_{26} - 2 \OpA_{34}
,\\
0 &= \OpA_{24} + \OpA_{26} - \OpA_{32} - \OpA_{34}
,\\
0 &= \OpA_{12} + 2 \OpA_{19} - 2 \OpA_{35}
,\\
0 &= \OpA_{13} + \OpA_{15} + \OpA_{17} + 2 \OpA_{27} - 2 \OpA_{37}
,\\
0 &= \OpA_{14} + \OpA_{16} + \OpA_{18} - 2 \OpA_{28} - 2 \OpA_{38}
,\\
0 &= \OpA_{21} + 2 \OpA_{29} - 2 \OpA_{39}
,\\
0 &= \OpA_{22} + \OpA_{23} + \OpA_{25} + \OpA_{31} + \OpA_{33} - 2 \OpA_{40}
.
\end{split}
\end{equation}
Likewise, for the P odd operators one finds the relations
\begin{equation}
\begin{split}
0 &= 2 \OpH_{1} - \OpH_{5} + \OpH_{11} - \OpH_{13} - 2 \OpH_{18}
,\\
0 &= 2 \OpH_{1} - \OpH_{2} + \OpH_{4} + \OpH_{5} - \frac{1}{2} \OpH_{10} - \OpH_{22}
,\\
0 &= \OpH_{5} - 2 \OpH_{7} + \OpH_{11} + \OpH_{13} - 2 \OpH_{24}
,\\
0 &= \OpH_{14} + \OpH_{17} - \OpH_{23} + 2 \OpH_{25}
.
\end{split}
\end{equation}

In addition, the P odd operators are not independent due to the
four-dimensional identity
\begin{equation}
X_{\mu, a, a, a, a} - X_{a, \mu, a, a, a} + X_{a, a, \mu, a, a} 
- X_{a, a, a, \mu, a} + X_{a, a, a, a, \mu} = 0
.
\end{equation}
As a consequence, the following relations exist
\begin{equation}
\begin{split}
0 &=  -\OpH_{1} + \OpH_{2} - \OpH_{3} - \OpH_{5} + \OpH_{6} - \OpH_{7}
,\\
0 &=
 -2 \OpH_{1} + \OpH_{2} - \OpH_{4} - \OpH_{5} + \OpH_{8}
,\\
0 &=
 -\OpH_{1} + \OpH_{3} - \OpH_{4} - \OpH_{5} - \OpH_{7} + \OpH_{9}
,\\
0 &=
 -\frac{1}{2} \OpH_{10} - \OpH_{11} + \OpH_{12}
,\\
0 &=
 \OpH_{15} - \OpH_{17} + \OpH_{19} - \OpH_{21} - \OpH_{23}
,\\
0 &=
 \OpH_{16} - \OpH_{18} + \OpH_{20} + \OpH_{22} - \OpH_{24}
.
\end{split}
\end{equation}

The effective actions $\Gamma_{4+2,a}^\pm$ and  $\Gamma_{4+2,b}^\pm$ for Majorana
neutrinos and $\Gamma_{4+2,D}^\pm$ for Dirac ones can be expressed in the form
\begin{equation}
\begin{split}
\Gamma_{4+2,a}^+ &= \int d^4 x \frac{v^2}{\phi^2} \sum_k i g^+_{a,k} A^+_k 
,\qquad
\Gamma_{4+2,a}^- = \int d^4 x \frac{v^2}{\phi^2}\sum_k i g^-_{a,k} A^-_k 
,\\
\Gamma_{4+2,b}^+ &= \int d^4 x \frac{v^2}{\phi^2} \sum_k i g^+_{b,k} A^+_k 
,\qquad
\Gamma_{4+2,b}^- = \int d^4 x \frac{v^2}{\phi^2} \sum_k i g^-_{b,k} A^-_k 
,\\
\Gamma_{4+2,D}^+ &= \int d^4 x \frac{v^2}{\phi^2} \sum_k i g^+_{D,k} A^+_k  
,\qquad
\Gamma_{4+2,D}^- = 0
.
\end{split}
\label{eq:6.21}
\end{equation}
The explicit imaginary unit has been introduced so that the couplings
$g^\pm_{t,k}$ are all real. The operators $A^\pm_k$ themselves are common for
the two structures $\Gamma_a$ and $\Gamma_b$, but the couplings are sensitive
to this structure.

\begin{table}[t]
\caption{Non vanishing couplings for $\Gamma_{4+2,a}^+$.}
\centering
\begin{tabular}{|lr|lr|lr|lr|}
\hline
 $ \gA_{a,1} \EQ $ & $ -\frac{1}{4} \IaA $ 
&
 $ \gA_{a,10} \EQ $ & $ -\frac{1}{6} \IaA $
&
 $ \gA_{a,20} \EQ $ & $ \frac{1}{2} \IaA -\frac{5}{3} \IaB $
&
 $ \gA_{a,33} \EQ $ & $ -2 \IaA $
\\
 $ \gA_{a,2} \EQ $ & $ \frac{5}{6} \IaA $ 
&
 $ \gA_{a,11} \EQ $ & $ -\frac{2}{3} \IaA $
&
 $ \gA_{a,22} \EQ $ & $ -\frac{2}{3} \IaA $
&
 $ \gA_{a,34} \EQ $ & $ \frac{1}{3} \IaA -\frac{2}{3} \IaB $
\\
 $ \gA_{a,3} \EQ $ & $ -\frac{1}{12} \IaA $ 
&
 $ \gA_{a,12} \EQ $ & $ \frac{1}{6} \IaA $
&
 $ \gA_{a,24} \EQ $ & $ \IaA -\frac{2}{3} \IaB $
&
 $ \gA_{a,35} \EQ $ & $ \frac{20}{3} \IaA + \frac{2}{3} \IaB $
\\
 $ \gA_{a,5} \EQ $ & $ -\frac{1}{3} \IaA $ 
&
 $ \gA_{a,13} \EQ $ & $ \frac{1}{3} \IaA $
&
 $ \gA_{a,26} \EQ $ & $ -\frac{1}{3} \IaA -\frac{2}{3} \IaB $
&
 $ \gA_{a,36} \EQ $ & $ 4 \IaA $
\\
 $ \gA_{a,6} \EQ $ & $ \frac{4}{3} \IaA $ 
&
 $ \gA_{a,15} \EQ $ & $ 2 \IaA $
&
 $ \gA_{a,28} \EQ $ & $ -\IaA + \frac{2}{3} \IaB $
&
 $ \gA_{a,37} \EQ $ & $ \frac{4}{3} \IaB $
\\
 $ \gA_{a,7} \EQ $ & $ - \IaA $ 
&
 $ \gA_{a,16} \EQ $ & $ \frac{5}{3} \IaA + \frac{1}{3} \IaB $
&
 $ \gA_{a,30} \EQ $ & $ -\IaA + 2\IaB $
&
 $ \gA_{a,38} \EQ $ & $ 4 \IaA $
\\
 $ \gA_{a,8} \EQ $ & $ \frac{1}{6} \IaA $
&
 $ \gA_{a,17} \EQ $ & $ -2 \IaA $
&
 $ \gA_{a,31} \EQ $ & $ 2 \IaA $
&
 $ \gA_{a,39} \EQ $ & $ -\frac{8}{3} \IaA $
\\
 $ \gA_{a,9} \EQ $ & $ \frac{2}{3} \IaA $
&
 $ \gA_{a,18} \EQ $ & $ -\frac{8}{3} \IaA + \frac{1}{3} \IaB $
&
 $ \gA_{a,32} \EQ $ & $ 3\IaA -\frac{2}{3} \IaB $
&
 $ \gA_{a,40} \EQ $ & $ -\frac{8}{3} \IaB $
\\
\hline
\end{tabular}\\
\label{tab:4}
\end{table}

\begin{table}[t]
\caption{Non vanishing couplings for $\Gamma_{4+2,b}^+$.}
\centering
\begin{tabular}{|lr|lr|lr|}
\hline
 $ \gA_{b,2} \EQ $ & $ -\IbA + \IbB $ & 
 $ \gA_{b,16} \EQ $ & $ -\frac{5}{2} \IbA + 4\IbB $ &
 $ \gA_{b,28} \EQ $ & $ 4 \IbA $ 
\\
 $ \gA_{b,8} \EQ $ & $ \IbA $ & 
 $ \gA_{b,17} \EQ $ & $ \frac{7}{2} \IbA - 3\IbB $ & 
 $ \gA_{b,35} \EQ $ & $ 10\IbA - 15\IbC + 2\IbD - \IbE $
\\
 $ \gA_{b,10} \EQ $ & $ -1 \IbA $ &
 $ \gA_{b,18} \EQ $ & $ \frac{5}{2} \IbA - 4\IbB $ &
 $ \gA_{b,36} \EQ $ & $ -4 \IbC $ 
\\
 $ \gA_{b,12} \EQ $ & $ \frac{3}{4} \IbA + \frac{1}{2} \IbB $ &
 $ \gA_{b,20} \EQ $ & $ \frac{13}{2} \IbA - 4\IbB - 2\IbD + \IbE $ &
 $ \gA_{b,37} \EQ $ & $ -8\IbA + 12\IbB $ 
\\
 $ \gA_{b,15} \EQ $ & $ -\frac{7}{2} \IbA + 3\IbB $ & 
 $ \gA_{b,27} \EQ $ & $ -4 \IbA $ & 
 $ \gA_{b,38} \EQ $ & $ -2\IbA + 8\IbB $ \\
\hline
\end{tabular}\\
\label{tab:5}
\end{table}

\begin{table}[t]
\caption{Non vanishing couplings for $\Gamma_{4+2,a}^-$.}
\centering
\begin{tabular}{|lr|lr|lr|lr|}
\hline
 $ \gH_{a,14} \EQ $ & $ \frac{2}{3} \IaA $
&
 $ \gH_{a,16} \EQ $ & $ 2 \IaA $
&
 $ \gH_{a,21} \EQ $ & $ -\frac{1}{3} \IaA $
&
 $ \gH_{a,24} \EQ $ & $ -2 \IaA $
\\
 $ \gH_{a,15} \EQ $ & $ -\frac{2}{3} \IaA $
&
 $ \gH_{a,18} \EQ $ & $ -\frac{2}{3} \IaA $
&
 $ \gH_{a,22} \EQ $ & $ \frac{5}{3} \IaA $
&
 $ \gH_{a,25} \EQ $ & $ -\frac{8}{3} \IaB $
\\ 
\hline
\end{tabular}\\
\label{tab:6}
\end{table}

\begin{table}[t]
\caption{Non vanishing couplings for $\Gamma_{4+2,b}^-$.}
\centering
\begin{tabular}{|lr|lr|lr|}
\hline
 $ \gH_{b,10} \EQ $ & $ \frac{1}{2} \IbA $
&
 $ \gH_{b,21} \EQ $ & $ \frac{1}{2} \IbA - \IbB $
&
 $ \gH_{b,22} \EQ $ & $ -\frac{1}{2} \IbA $
\\
\hline
\end{tabular}\\
\label{tab:7}
\end{table}

\begin{table}[h]
\caption{Non vanishing couplings for $\Gamma_{4+2,D}^+$.}
\centering
\begin{tabular}{|lr|lr|lr|lr|}
\hline
 $ \gA_{b,1} \EQ $ & $ -\frac{1}{6} \IaA $ & 
 $ \gA_{b,15} \EQ $ & $ \frac{4}{3} \IaA $ & 
 $ \gA_{b,30} \EQ $ & $ \frac{4}{3} \IaA $ & 
 $ \gA_{b,37} \EQ $ & $ \frac{4}{3} \IaA $  
\\
 $ \gA_{b,2} \EQ $ & $ \frac{5}{6} \IaA $ & 
 $ \gA_{b,16} \EQ $ & $ \frac{4}{3} \IaA $ & 
 $ \gA_{b,31} \EQ $ & $ \frac{4}{3} \IaA $ &
 $ \gA_{b,38} \EQ $ & $ 2 \IaA $ 
\\
 $ \gA_{b,3} \EQ $ & $ -\frac{1}{6} \IaA $ & 
 $ \gA_{b,17} \EQ $ & $ -\frac{4}{3} \IaA $ & 
 $ \gA_{b,32} \EQ $ & $ \frac{4}{3} \IaA $ & 
 $ \gA_{b,39} \EQ $ & $ -\frac{8}{3} \IaA $  
\\
 $ \gA_{b,4} \EQ $ & $ \frac{2}{3} \IaA $ &
 $ \gA_{b,18} \EQ $ & $ -\frac{4}{3} \IaA $ & 
 $ \gA_{b,33} \EQ $ & $ -\frac{4}{3} \IaA $ & 
 $ \gA_{b,40} \EQ $ & $ -\frac{8}{3} \IaA $ 
\\
 $ \gA_{b,5} \EQ $ & $ -\frac{1}{3} \IaA $ & 
 $ \gA_{b,19} \EQ $ & $ -\frac{4}{3} \IaA $ &
 $ \gA_{b,34} \EQ $ & $ -\frac{4}{3} \IaA $ &
& 
\\
 $ \gA_{b,6} \EQ $ & $ \frac{2}{3} \IaA $ & 
 $ \gA_{b,20} \EQ $ & $ -\frac{4}{3} \IaA $ & 
 $ \gA_{b,35} \EQ $ & $ \frac{16}{3} \IaA $ &
&
\\
 $ \gA_{b,7} \EQ $ & $ -1 \IaA $ & 
 $ \gA_{b,29} \EQ $ & $ \frac{4}{3} \IaA $ & 
 $ \gA_{b,36} \EQ $ & $ 2 \IaA $ &
&
 \\
\hline
\end{tabular}\\
\label{tab:8}
\end{table}

The non vanishing couplings are collected in Tables
\ref{tab:4}, \ref{tab:5}, \ref{tab:6}, \ref{tab:7}, and \ref{tab:8}.
The couplings are expressed in terms of a few independent momentum integrals
(defined in \neq{4.8}),
namely,
\begin{equation}
\begin{split}
\IaA &\equiv I^6_{a, 1, 1, 2, 2}, \quad
\IaB \equiv I^8_{a, 1, 1, 2, 3},  \quad
\IbA \equiv I^2_{b, 1, 1, 1, 2}, \quad
\IbB \equiv I^4_{b, 1, 1, 1, 3}, \quad
\\
\IbC &\equiv I^4_{b, 1, 1, 2, 2}, \quad
\IbD \equiv I^6_{b, 1, 1, 3, 2}, \quad
\IbE \equiv I^6_{b, 2, 1, 2, 2}
.
\end{split}
\label{eq:6.22}
\end{equation}
In \Eq{6.21} we have extracted the main dependence on (underivated) $\phi$
from the couplings assuming a Dirac-type Higgs coupling in mass
terms. Therefore, in the various momentum integrals above $m_e$ no longer
contains the factor $(\phi/v)$ and $m_\nu$ contains a single factor $(\phi/v)$
in the Majorana case and none in the Dirac case.

It should be noted that not all the integrals $I_a$ and $I_b$ are
independent. The following relations have been used to simplify the
expressions:
\begin{equation}
\begin{split}
5 I^6_{a,1,1,2,2} &=  2 I^8_{a,1,1,2,3} + 2 I^8_{a,1,1,3,2}
\,,\\ 
 3 I^2_{b,1,1,1,2} &= 2 I^4_{b,1,1,1,3} + 2 I^4_{b,1,1,2,2}
\,,\\
 4 I^4_{b, 1, 1, 2, 2} &= 2 I^6_{b, 1, 1, 2, 3} + 2 I^6_{b, 1, 1, 3, 2} + I^6_{b, 2, 1, 2, 2}
\,, \\
 4 I^4_{b, 1, 1, 1, 3} &= 3 I^6_{b, 1, 1, 1, 4} + 2 I^6_{b, 1, 1, 2, 3} + I^6_{b, 1, 2, 1, 3}
\,.
\end{split}
\end{equation}
These relations follow from integration by parts in momentum space and the
symmetry or antisymmetry properties of $I_a$ and $I_b$ (more detailed
information, such as $U$ being a unitary matrix in $N_e$, is not required).

\subsection{\textsf{Discussion of the analytical results}}

The expressions given for the effective action are in Euclidean space.  With the
conventions of \cite{GarciaRecio:2009zp}, the expressions in Minkowskian space
take exactly the same form except for the two following modifications:
$\epsilon_{\mu\nu\alpha\beta} \to i \epsilon_{\mu\nu\alpha\beta}$ (in
$\Gamma^-$) and $Z_\mu \to i Z_\mu$ (as well as $Z_{\mu\nu} \to i Z_{\mu\nu}$,
etc). The meaning of the symbols changes to conform to the Minkowskian
conventions, and so $(W^+)^*_\mu= W^-_\mu$ and $Z_\mu$, $\varphi_\mu$ and
$F^e_{\mu\nu}$ are real. The resulting real-time effective action is real,
both for the parity even and odd components. The true effective action (in
Minkowski space) has an imaginary part when the vertex functions contained in
it are above the unitarity thresholds. In our calculation we are always below
thresholds since the derivative expansion is an expansion around zero
momentum.

The set of operators is common to the effective actions obtained from
integration of leptons or from integration of quarks. For each operator, the
total coupling is obtained by adding the quark and lepton contributions. Also,
the operators do not distinguish between the couplings of type $I_a$, which
corresponds to alternating charged bosons along the loop, $W^+W^-W^+W^-$, and
those of type $I_b$, with structure $W^+W^+W^-W^-$, which is exclusive for
Majorana neutrinos. The total coupling to a given operator is obtained by
adding its type $I_a$ and type $I_b$ contributions. Integrals of the two types
$I_a$ and $I_b$ appear in parity even and parity odd operators.

For Dirac neutrinos, as for quarks, all the couplings to P odd operators
vanish in the CP sector. The single more interesting result found here is that
Majorana neutrinos would produce C even P odd terms in $\Gamma_{4+2}$. This can
be obtained without recourse to $Z$ or Higgs fields, namely,
\begin{equation}
\frac{1}{2} i b_1 \epsilon_{\mu\nu\alpha\beta}( W^+_{\mu\nu} W^+_{\alpha\beta} 
W^-_\rho W^-_\rho + \text{\rm c.c.} )
\,.
\end{equation}
This term corresponds to $A^-_{10}$ and the coupling is purely of Majorana
type ($I_b$). Further operators of types $I_a$ and $I_b$ are allowed if $Z$
and $\varphi$ are included.\footnote{\cite{Hernandez:2008db} reported a non
  null coupling to $A_{21}^-$ for quarks, which would result in a similar
  coupling for Dirac neutrinos, however that coupling has been shown to vanish
  in \cite{GarciaRecio:2009zp,Brauner:2011vb,Salcedo:2011hy}. The calculations
  in \cite{GarciaRecio:2009zp,Brauner:2011vb} are based on \Eq{2.20}, first
  derived in \cite{Salcedo:2008tc}, and so their method differs from that used
  in the calculation of \cite{Hernandez:2008db}. On the other hand the
  calculations of \cite{Smit:2004kh,Hernandez:2008db,Salcedo:2011hy} follow
  the method given in \cite{Salcedo:2000hx} which uses the current to
  reconstruct the effective action.} It is also interesting that this term can
be used to show that CP would be violated, for generic values of $U$ and the
masses, in a two generations scenario. Of course, the same would not be true
for Dirac neutrinos or quarks, which require at least three generations.

An inspection of the results shows that $F^e_{\mu\nu}$ is not present. For P
even terms this follows from the fact that the only $4+2$ operator one can
write involving $F^e_{\mu\nu}$ is $F^e_{\mu\nu} W^+_\mu W^+_\alpha W^-_\nu
W^-_\alpha$, which is CP even. On the other hand, in the P odd sector the
unique operator with $F^e_{\mu\nu}$ is $A^-_{13}$ and it is CP odd, however, it
can be eliminated using integration by parts.

The effective action for Dirac neutrinos is formally identical to that
obtained for quarks in \cite{GarciaRecio:2009zp}. The calculation there was
carried out for generic gauge connections $A^u_\mu$ and $A^d_\mu$ (which in
the Standard Model take the values $A_u=-(2/3)A_e$ and $A_d=+(1/3)A_e$), and
so it includes the leptonic case for Dirac neutrinos by setting $A_u=0$
$A_d=A_e$. Since $F^u_{\mu\nu}$, $F^d_{\mu\nu}$ do not appear in the formulas,
the results are formally equal. Moreover, the derivatives of the charged gauge
fields are also equal since $A_d-A_u = A_e$. As said the final operators are
common to leptons and quarks. The difference between the CP violating
effective actions induced by quarks and Dirac leptons comes only from the
difference in the mass matrices (i.e., $U$ and masses). We dwell on this in
Sec. \ref{subsec:7.4}.

For Majorana neutrinos, the term $\Gamma^{\pm}_{4+2,b}$ is new and represents
a different mechanism which involves (virtual) lepton-number violation. The
terms $\Gamma^{\pm}_{4+2,a}$ use the same Kobayashi-Maskawa mechanism as
quarks or Dirac neutrinos. As already noticed $\Gamma_{4+2,a}$ coincides with
$\Gamma_{4+2,D}$ modulo terms involving $Z$ and Higgs. The explicit
calculation shows that they differ in terms depending on $Z$ or $\varphi$. In
particular, $\Gamma^-_{4+2,a}$ does not vanish (whereas $\Gamma^-_{4+2,D}=0$)
and it receives contributions from operators $A^-_k$ with $k\ge 14$ (operators
with $Z$ or $\varphi$).

The effective action functional for Dirac neutrinos has a number of
interesting regularities. They are more clearly exposed by writing
$\Gamma^+_{4+2,D}$ explicitly as done in Eqs. (10.1-2) of
\cite{GarciaRecio:2009zp} (for quarks). First, there is just one coupling,
$a_1$, for all the terms. More interestingly, $\varphi$ and $Z$ appear solely
in the form $\Im F[\varphi + Z] $ (recall that $(\varphi + Z)^* = (\varphi -
Z)$ in Euclidean space). This suggests some kind of analytical dependence
since clearly, this is not the most general possible dependence of a
functional on the two variables $\varphi$ and $Z$. A well defined pattern of
dependence on $Z \pm \varphi$ was identified at the level of $K_{\rm Dirac}$
in \neq{5.13}, as well as in $K_1$ in \neq{4.10} for Majorana neutrinos, but
the implications are not obvious since the number of $\varphi$'s is not
preserved by subsequent manipulations.

It is not clear why or how, a (complex) variation in $\varphi$ could be
canceled by doing a similar variation in the $Z$. We have been unable to
verify whether this ``symmetry'' persists in other terms of the effective
action (not necessarily of the type $4+2$ and CP odd), and it is possible that
this is just an accidental symmetry due to the low order of the terms
considered.  Indeed, at the level of $4+2$ not many terms can be written
violating the structure $\Im F[\varphi + Z]$, essentially only those of the
form $W^+W^+W^-W^-(\varphi + Z)(\varphi - Z)$.

We have investigated whether the structure $\Gamma \sim \Im F[\varphi + Z] $
also shows up in the effective action of the Majorana neutrinos,
$\Gamma_{4,M}$. From the results shown in the tables, we find that the
symmetry persists in $\Gamma_{4+2,a}^+$ and $\Gamma_{4+2,b}^-$. In
$\Gamma_{4+2,b}^+$ it is broken by a term $i2(b_1+b_2)(A^+_{38} - A^+_{36})$,
and in $\Gamma_{4+2,a}^-$ it is broken by a term $i(8/3)a_2
A^-_{25}$. Nevertheless, to reach a firm conclusion it would be necessary to
lift the simplifying assumption $\phi \, (\partial I / \partial\phi) = -2 I$
(see \neq{6.16}) in the integration by parts, which affects the dependence on
$\varphi_\mu$.

Another regularity found in $\Gamma^+_{4+2,D}$ is that it has the structure
\begin{equation}
\begin{split}
\Gamma^+_{4+2,D} &\sim F_0[W^+ W^+ \hat{D}W^-\hat{D}W^-] + 
F_1[W^+ W^+ W^-\hat{D}W^- (\varphi+Z)] 
\\
& ~~~
+ 
F_2[W^+ W^+ W^- W^- (\varphi+Z)(\varphi+Z)]
+ \text{\rm c.c.}
\end{split}
\end{equation}
The position of the derivatives is also not the most general one, even after
integration by parts. This suggests that the $(\varphi+Z)$ dependence could be
recovered from $F_0$ by some kind of gauging, $\hat D \to \hat D + \varphi+Z$,
but we have been unable to establish whether such a gauging exists.

\section{\textsf{Invariants and couplings}}
\label{sec:7}

In this section we analyze the dependence on $U$ and on the lepton masses of
the results just obtained and focus on the couplings of two concrete
paradigmatic cases. Throughout this section the diagonal mass matrices of
charged leptons and neutrinos are denoted $\hm_e$ and $\hm_\nu$, reserving
$m_e$ for the electron mass. Also, no factors of $\phi/v$ are implicit.

\subsection{\textsf{Invariants}}

The momentum integrals $I_a$ and $I_b$ contain two different structures in
flavor space
\begin{equation}
\begin{split}
I_a &\sim \Im \tr \left( 
U^\dagger f_1(\hm_e) \, U f_2(\hm_\nu)  \, U^\dagger f_3(\hm_e)
\, U 
f_4(\hm_\nu) \right)
,
\\
I_b &\sim \Im \tr \left(
U^\dagger f_1(\hm_e) \, U f_2(\hm_\nu)  \, U^T f_3(\hm_e) \,
U^* 
f_4(\hm_\nu) \right)
.
\end{split}
\label{eq:7.2}
\end{equation}
The first structure is common to Dirac and Majorana cases while the second
structure is specific for Majorana neutrinos. By expanding in matrix
elements, the $U$-dependent tensors relevant for $I_a$ and $I_b$ are found to
be, respectively,
\begin{equation}
J_{\alpha\beta}^{ij} \equiv \Im (Z^{ij}_\alpha Z^{ji}_\beta )
,
\qquad
K_{\alpha\beta}^{ij} \equiv \Im (Z^{ij}_\alpha Z^{ij}_\beta )
,
\end{equation}
where
\begin{equation}
Z^{ij}_\alpha  \equiv U_{\alpha i} U^*_{\alpha j}
= Z^{ji}_\alpha {}^*
.
\end{equation}
Following the standard practice, the labels $\alpha$, $\beta$, $\gamma$, etc,
refer to the charged leptons and $i$, $j$, $k$, etc, to the neutrinos with
well-defined masses \cite{Morii:2004tp,Beringer:1900zz}. All the algebraic
properties of the tensors $J$ and $K$ stem from the fact that $U$ is unitary,
and this information can be codified in $Z$ under the conditions that, as a
matrix with respect to $ij$, i) $Z_\alpha=Z_\alpha^\dagger$, ii) $Z_\alpha
Z_\beta = \delta_{\alpha\beta} Z_\alpha$, and $\tr(Z_\alpha)=1$ (i.e., the
three $Z_\alpha$ are orthogonal projectors on one-dimensional subspaces of
$\C^3$).  Unfortunately these necessary and sufficient conditions are not
linear.

The tensor $J$ is antisymmetric with respect to $ij$ and to $\alpha\beta$
and in fact, for $g=3$, it has only one independent component, the well-known
Jarlskog invariant \cite{Jarlskog:1985ht}
\begin{equation}
J^{ij}_{\alpha\beta} = J_{\rm CP}  \qquad \text{for~} (ijk) \text{~and~}
(\alpha\beta\gamma) \text{~cyclic}. 
\end{equation}
Because $Z^{ij}_\alpha$ is invariant under phase redefinitions of the charged
leptons, $U_{\alpha i} \to e^{i \varphi_\alpha} U_{\alpha i}$, so are
$J^{ij}_{\alpha\beta}$ and $K^{ij}_{\alpha\beta}$. The symmetry of
$J^{ij}_{\alpha\beta}$ is larger since it is also invariant under phase
redefinitions of the neutrino fields, $U_{\alpha i} \to U_{\alpha i}
e^{i \varphi_i}$. This is also the situation for quarks.

Remarkably, although the charged leptons are Dirac fermions as the quarks, the
matrix specifically relevant for Majorana neutrinos, $K^{ij}_{\alpha\beta}$
is still antisymmetric in the neutrino sector $ij$ but symmetric in the
charged lepton sector, $\alpha\beta$,
\begin{equation}
K^{ij}_{\alpha\beta} = -K^{ji}_{\alpha\beta} = K^{ij}_{\beta\alpha}
.
\end{equation}
In principle this reduces the independent components in $K$ from $81$ to $18$
(for $g=3$). However, the property $\sum_\alpha Z^{ij}_\alpha = \delta_{ij}$,
implies the further 9 conditions
\begin{equation}
\sum_\alpha K^{ij}_{\alpha\beta} = 0
,
\end{equation}
which leaves just $9$ linearly independent components in
$K^{ij}_{\alpha\beta}$ for $g=3$. There are no further linear constraints.

For two generations $J^{ij}_{\alpha\beta}$ vanishes identically (hence the
need of at least three flavors to break CP \cite{Kobayashi:1973fv}) but
$K^{ij}_{\alpha\beta}$ has still one non null component. As a consequence CP
violation is allowed in the two generations version of the Standard Model
minimally extended to include Majorana neutrinos \cite{Morii:2004tp}.  We have
verified that this is actually the case in our calculation, i.e., there are no
accidental cancellations, and so for instance, the coupling to $A_{10}^-$ is
not zero for generic $2\times 2$ unitary $U$ and generic lepton masses.

Coming back to three generations, the tensor $K^{ij}_{\alpha\beta}$,
being antisymmetric in $ij$, can be arranged into three symmetric matrices
with respect to $\alpha\beta$, a matrix for each cyclic $(ij)$. Moreover, the
sum by columns or by rows in these matrices vanishes and this allows to use
the cyclic $(\alpha\beta)$ components to parameterize them:
\begin{equation}
K^{ij} = \begin{pmatrix}
-K^k_\mu - K^k_\tau & K^k_\tau & K^k_\mu \\
K^k_\tau & -K^k_e - K^k_\tau & K^k_e \\
K^k_\mu &  K^k_e & -  K^k_e - K^k_\mu
\end{pmatrix}
,
\quad
K^k_\gamma \equiv K^{ij}_{\alpha\beta}
\text{~~(cyclic~}
(ijk) \text{~and~}
(\alpha\beta\gamma)
)
.
\end{equation}
(For three generations the diagonal matrix elements $K^{ij}_{\alpha\alpha}$
also serve as independent parameters.)

The invariance under phase redefinitions of the charged leptons removes
(renders ineffective) three out of the nine parameters in $U$, leaving only 6
effective parameters in $K^{ij}_{\alpha\beta}$, namely, 3 angles, one Dirac
phase and two Majorana phases. (We have verified that the 6 parameters are
truly effective, i.e., the $K^k_\gamma$ fill a six-dimensional submanifold of
$\R^9$ as $U$ moves in SU(3).) This implies that not all the nine linearly
independent components $K^k_\gamma$ are truly algebraically independent.

In order to find new constraints, we note that, for any $(ij)$, the
determinant of the matrix $K^{ij}$ vanishes while the determinants of the
three $2\times 2$ submatrices are all equal. What is not so trivial is that
these determinants are actually independent of the label $(ij)$. Indeed,
\begin{equation}
K^{ij}_{\alpha\alpha} K^{ij}_{\beta\beta}
-
K^{ij}_{\alpha\beta} K^{ij}_{\beta\alpha}
=
J^{ij}_{\alpha\beta} J^{ij}_{\beta\alpha}
=
-J_{\rm CP}^2
  \qquad \text{for~} i\not=j \text{~and~}
\alpha\not=\beta . 
\end{equation}
The first equality follows just from the definitions of the tensors $K$ and
$J$ in terms of $Z$ in (\ref{eq:7.2}), while the second equality requires
$i\not=j$ and $\alpha\not=\beta$ and relies on the fact that there is just a
single independent component in $J^{ij}_{\alpha\beta}$. More explictly,
\begin{equation}
K^k_e K^k_\mu + K^k_\mu K^k_\tau + K^k_\tau K^k_e + J_{\rm CP}^2 = 0
,\qquad
k=1,2,3
.
\label{eq:7.9}
\end{equation}
This relation implies that the nine invariants $K^k_\gamma$ can be expressed
in terms of six of them plus the Jarlskog invariant, or equivalently, in terms
of seven of them. The number of independent parameters is six. This suggests
that there exist a further non linear relation among the nine invariants,
presumably of polynomial type, but we have not found it.

It is interesting that a $2\times 2$ symmetric submatrix can be identified
with a bidimensional metric and so with a ellipse (the three parameters being
the two principal lengths and one rotation angle). Each of the three $K^{ij}$
is equivalent to one such ellipse, and the identity in (\ref{eq:7.9}) implies
that they have the same area. It can be speculated that the missing constraint
is related to some other geometrical property of these figures.

\subsection{\textsf{Couplings of Dirac type}}

In order to analyze the couplings obtained, as regards to CP violation, we
consider two cases, one of type $I_a$, common to Dirac and Majorana neutrinos,
and another of type $I_b$ for Majorana neutrinos.

For Dirac neutrinos, all the couplings are proportional to $a_1$ so we
consider this case. The same coupling appears also for Majorana neutrinos.
\begin{equation}
a_1 = I^6_{a,1,1,2,2} = \Im\int \frac{d^4p}{(2\pi)^4} p^6 \tr( U^\dagger \hn_e U
\hn_\nu U^\dagger \hn_e^2 U \hn_\nu^2 )
\end{equation}
where the propagators $\hn_{e,\nu}$ are diagonal matrices
\begin{equation}
\hn_e = (p^2 + \hm_e^2)^{-1} 
,\qquad
\hn_\nu = (p^2 + \hm_\nu^2)^{-1} 
.
\end{equation}
This integral is identical to that for quarks in \cite{GarciaRecio:2009zp}, so
the results can be taken from there in a direct way:
\begin{equation}
a_1 = J_{\rm CP} \Delta_\nu \Delta_e I_{\nu e}
,
\label{eq:7.12}
\end{equation}
where
\begin{equation}
\begin{split}
\Delta_\nu &= (m_{\nu,1}^2-m_{\nu,2}^2) (m_{\nu,2}^2-m_{\nu,3}^2) (m_{\nu,3}^2-m_{\nu1}^2)
,
\\
\Delta_e &= (m_e^2-m_\mu^2) (m_\mu^2-m_\tau^2) (m_\tau^2-m_e^2)
,
\\
I_{\nu e} &= 
\int \frac{d^4p}{(2\pi)^4} p^6 \prod_{i=1}^3 N_{\nu,i}^2 
\prod_{\alpha=1}^3 N_{e,\alpha}^2
. 
\end{split}
\end{equation}

At this point approximations can be taken exploiting the big difference
between the mass scales of neutrinos and charged leptons. For a generic
momentum integral with heavy and light masses
\begin{equation}
I_{lh} = \int \frac{d^4p}{(2\pi)^4} p^{n-4} \prod_l N_l^{n_l} \prod_h N_h^{n_h}
\qquad
( 0 < n < 2 \sum_l n_l + 2 \sum_h n_h )
\label{eq:7.14}
\end{equation}
the effective integration range of the variable $p$ is fixed by the light
masses and the momentum can be neglected in the heavy propagators, provided the
remaining integral is still UV convergent,
\begin{equation}
\begin{split}
I_{lh} &=   \frac{1}{\prod_h m_h^{2n_h}} I_l
\times
\left(1+O\left({\bar{m}_l^2}/{\bar{m}_h^2}\right)\right)
,\\
I_l &= 
\int \frac{d^4p}{(2\pi)^4} p^{n-4} \prod_l N_l^{n_l} 
\qquad
(n < 2 \sum_l n_l)
.
\end{split}
\label{eq:7.15}
\end{equation}

For the coupling $a_1$ this implies
\begin{equation}
a_1 \approx J_{\rm CP} \hat\Delta_e \, \Delta_\nu I_\nu
,
\end{equation}
where
\begin{equation}
\begin{split}
\hat\Delta_e
&= 
\left( \frac{1}{m_\mu^2} -  \frac{1}{m_e^2} \right)
\left( \frac{1}{m_\tau^2} -  \frac{1}{m_\mu^2} \right)
\left( \frac{1}{m_e^2} -  \frac{1}{m_\tau^2} \right)
\approx
\frac{1}{m_e^4 m_\mu^2}
,
\\
I_\nu  &= 
\int \frac{d^4p}{(2\pi)^4} p^6 N_{\nu,1}^2  N_{\nu,2}^2  N_{\nu,3}^2 
\end{split}
\end{equation}

The experimental value of the leptonic invariant $J_{\rm CP}$ is not yet
well determined, since the value of the Dirac phase is not known. From its
definition $|J_{\rm CP}| \le 1/(6\sqrt{3}) = 0.096$ and current data on the
angles imply $|J_{\rm CP}| < 0.039$ \cite{Beringer:1900zz}.

The correct hierarchy of masses, namely, $m_{\nu,1} < m_{\nu,3}$ (normal
hierarchy) or $m_{\nu,1} > m_{\nu,3}$ (inverted hierarchy) is not yet
known.\footnote{ We adopt the standard choice for labeling the neutrinos,
  namely,
$ m_{\nu,1} < m_{\nu,2} $, and 
$m_{\nu,2} - m_{\nu,1} < \min(|m_{\nu,3} - m_{\nu,1}|,|m_{\nu,3} - m_{\nu,2}|)$.
}
The data on differences between square masses are currently becoming rather
precise from several neutrino oscillation experiments
\cite{Ahn:2012nd,Abe:2012tg,Abe:2013xua,An:2013uza}. With the usual notation,
$\Delta m_{ij}^2 = m_{\nu,i}^2 - m_{\nu,j}^2$, the data indicate that 
$\Delta m_{21}^2 \ll |\Delta m_{31}^2|$, and so
\begin{equation}
\Delta_\nu \approx  \Delta m_{21}^2 |\Delta m_{31}^2|^2
.
\end{equation}
Specifically, $\Delta m_{21}^2 = (8.7\pm 0.2 \, {\rm meV})^2$ and $|\Delta
m_{31}^2| = (49\pm 1 \, {\rm meV})^2$ ~\cite{Beringer:1900zz}.\footnote{Note
  that $|\Delta m_{ij}|^2 \le |\Delta m_{ij}^2|$, the equal sign requiring a
  massless neutrino.}

On the other hand, there is no precise information on the absolute values of
the masses, although the situation is rapidly changing for upper bounds on the
sum of neutrino masses, $m_{\nu,1}+m_{\nu,2}+m_{\nu,3}$, from astrophysics
data analyzed using available cosmological models
\cite{Abazajian:2011dt,Ade:2013zuv}. These bounds are in the range
$(0.3-1.3)\,{\rm eV}$, depending on the data and model used
\cite{Beringer:1900zz}.

Rather than computing $I_\nu$ for generic values of the masses, we will
consider three typical scenarios.

a) \textbf{Quasi degenerate.} $m_{\nu,i} \approx \bar{m}_\nu \gg |\Delta
m_{31}^2|^{1/2}$ ($\bar{m}_\nu$ is a fraction of eV). Using the relation
\begin{equation}
\int \frac{d^4p}{(2\pi)^4} \frac{p^{2n-4}}{(p^2 + m^2)^s} 
=
\frac{1}{(4\pi)^2} 
\frac{\Gamma(n)\Gamma(s-n)}{\Gamma(s)}
\frac{1}{m^{2(s-n)}}
\qquad
(0 < n < s)
,
\label{eq:7.19}
\end{equation}
one obtains in this case
\begin{equation}
\Delta_\nu I_\nu \approx  \frac{1}{5} \frac{1}{(4\pi)^2} \frac{1}{\bar{m}_\nu^2}
 \Delta m_{21}^2 |\Delta m_{31}^2|^2
.
\label{eq:7.20}
\end{equation}

b) \textbf{Normal hierarchy.} $m_{\nu,1} \ll m_{\nu,2} \approx (\Delta
m_{21}^2)^{1/2}$, and $m_{\nu,3} \approx |\Delta m_{31}^2|^{1/2}$.  In this
case the integral $I_\nu$ is of the type in (\ref{eq:7.14}), with $m_{\nu,1}$
and $m_{\nu,2}$ as the light masses and $m_{\nu,3}$ as the heavy one.  An
estimate of $I_{lh}$ can be obtained by neglecting the light masses in the
propagator, provided the remaining integral is still IR convergent,
\begin{equation}
\begin{split}
I_{lh} &=   I_h
\times
\left(1+O\left({\bar{m}_l^2}/{\bar{m}_h^2}\right)\right)
,\\
I_h &= 
\int \frac{d^4p}{(2\pi)^4} p^{n-4-2\sum_l n_l} \prod_h N_h^{n_h}
\qquad
(n > 2 \sum_l n_l)
.
\end{split}
\label{eq:7.21}
\end{equation}
In our case (using (\ref{eq:7.19}))
\begin{equation}
\Delta_\nu I_\nu \approx \frac{1}{(4\pi)^2}\Delta m_{21}^2 |\Delta m_{31}^2|
.
\end{equation}

c) \textbf{Inverted hierarchy.} $m_{\nu,3} \ll m_{\nu,1} \approx m_{\nu,2}
\approx  (\Delta m_{31}^2)^{1/2}$. In this case $m_{\nu,3}$ is light and
$m_{\nu,1}$ and $m_{\nu,2}$ are heavy in $I_\nu$ and we can apply
(\ref{eq:7.21}). This gives
\begin{equation}
\Delta_\nu I_\nu \approx \frac{1}{3} \frac{1}{(4\pi)^2} 
\Delta m_{21}^2 |\Delta m_{31}^2|
.
\end{equation}

\subsection{\textsf{Couplings of Majorana type}}

Here we consider the coupling $b_1$, typical of $I_b$ type
\begin{equation}
\begin{split}
b_1 &= I^2_{b,1,1,1,2}
=
\Im \int \frac{d^4p}{(2\pi)^4} p^2
\tr( U^\dagger \hn_e U \hm_\nu \hn_\nu U^T \hn_e U^* \hm_\nu \hn_\nu^2 )
\\
&=
\sum_{i,j,\alpha,\beta} K^{ij}_{\alpha\beta} \, 
 m_{\nu,i} m_{\nu,j} \int \frac{d^4p}{(2\pi)^4} p^2 N_{\nu,i} N_{\nu,j}^2
N_{e,\alpha} N_{e,\beta}
.
\end{split}
\end{equation}

At this point, we recall that $K^{ij}_{\alpha\beta}$ contains just nine
linearly independent components through antisymmetry in $ij$, and in addition
\begin{equation}
K^{ij}_{\gamma\gamma} = 
- K^{ij}_{\gamma\beta} - 
K^{ij}_{\gamma\alpha} 
=  - K^k_\alpha - K^k_\beta 
\qquad
(\text{cyclic~}
(ijk) \text{~and~}
(\alpha\beta\gamma)
)
\end{equation}
Using the identities
\begin{equation}
\begin{split}
N_{\nu,i} N_{\nu,j}^2 - N_{\nu,j} N_{\nu,i}^2
& = (m_{\nu,i}^2 - m_{\nu,j}^2) N_{\nu,i}^2 N_{\nu,j}^2,
,
\\
2 N_{e,\alpha} N_{e,\beta}
-  N_{e,\alpha}^2 -  N_{e,\beta}^2
&=
- (m_{e,\alpha}^2 - m_{e,\beta}^2)^2 N_{e,\alpha}^2 N_{e,\beta}^2 
,
\end{split}
\end{equation}
the coupling can be written as
\begin{equation}
b_1 = -
\sum_{k,\gamma} K^k_\gamma 
m_{\nu,i} m_{\nu,j} (m_{\nu,i}^2 - m_{\nu,j}^2) (m_{e,\alpha}^2 - m_{e,\beta}^2)^2
\int \frac{d^4p}{(2\pi)^4} p^2 N_{\nu,i}^2 N_{\nu,j}^2 N_{e,\alpha}^2
N_{e,\beta}^2
. 
\end{equation}
In this formula $(ijk)$ and $(\alpha\beta\gamma)$ are cyclic permutations of
$(123)$ and $(e\mu\tau)$, respectively.

The expression contains nine integrals. For generic masses one could reduce
this number to five by adding in each case the missing neutrino or charged
lepton propagator squared with $1=N^2 (p^2+m^2)^2$. In the case at hand, it is
preferable to exploit that the charged leptons are much heavier than the
neutrinos to factorize the expression, using (\ref{eq:7.15}). This gives,
\begin{equation}
b_1 = -
\sum_{k,\gamma} K^k_\gamma
F_{e,\gamma} F_{\nu,k}
\times 
\left(1+O\left({m_\nu^2}/{m_e^2}\right)\right)
\approx -\sum_k f_k F_{\nu,k}
,
\end{equation}
with
\begin{equation}
\begin{split}
F_{e,\gamma} &= \left(\frac{1}{m_{e,\alpha}^2} -  \frac{1}{m_{e,\beta}^2} \right)^2
,
\qquad
f_k = \sum_\gamma K^k_\gamma F_{e,\gamma} \approx
\frac{K^k_e}{m_\mu^4} + \frac{K^k_\mu}{m_e^4} + \frac{K^k_\tau}{m_e^4}
,
\\
F_{\nu,k} &= 
m_{\nu,i} m_{\nu,j} \Delta m^2_{ij}
\int \frac{d^4p}{(2\pi)^4} p^2 N_{\nu,i}^2 N_{\nu,j}^2
\equiv \frac{1}{(4\pi)^2}  \Delta m^2_{ij} H(m_{\nu,i}^2/m_{\nu,j}^2)
.
\end{split}
\end{equation}
Here we have introduced the dimensionless function
\begin{equation}
H(x) = (4\pi)^2 m_1 m_2 \int \frac{d^4p}{(2\pi)^4} p^2 
\frac{1}{(p^2+m_1^2)^2}\frac{1}{(p^2+m_2^2)^2}
,\qquad
x = \frac{m_1^2}{m_2^2}
.
\end{equation}
The integral can be conveniently obtained by residues (using
e.g. Eqs.~(11.6,7) of \cite{GarciaRecio:2009zp})
\begin{equation}
H(x) = \sqrt{x}
\frac{x^2 - 1 - 2 x \log x}{(x-1)^3}
,
\qquad
x>0
.
\end{equation}
This function increases monotonically from $x=0$ to $x=1$, further
\begin{equation}
H(x) = H(1/x)
,\quad
H(x) = \sqrt{x} + O(x^{3/2} \log x)
 ,\quad
H(1) = \frac{1}{3}
.
\end{equation}

At present nothing is known about the phases in $U$ and so only bounds can be
given on the nine invariants $K^k_\gamma$. Due to symmetry in the labels, the
maximum value of each $|K^k_\gamma|$, namely $1/4$, is common to all the
invariants (but, of course, this extreme is not attained simultaneously for
all of them). For instance, for $K^2_\mu$ it is attained for
$\theta_{13}=\pi/4$, $\alpha_{13}=\pi/2$ and
$\theta_{12}=\theta_{23}=\alpha_{12}=\delta_{\rm CP}=0$ (among other sets of
values). Here we have used a standard notation for the parameters in $U$
\cite{Beringer:1900zz}.

To estimate the effect of the neutrino factor $F_{\nu,k}$ we take the same
three scenarios considered previously for Dirac type integrals:

a) \textbf{Quasi degenerate.} If the neutrino masses are similar, $H(1)=1/3$
applies and this gives
\begin{equation}
b_1 \approx -\frac{1}{3}\frac{1}{(4\pi)^2} 
\left( (f_1-f_3) \Delta m^2_{21} + (f_2-f_1) \Delta m^2_{31} \right)
.
\end{equation}

b) \textbf{Normal hierarchy.} In this case $m_{\nu,1} \ll m_{\nu,2} \ll
m_{\nu,3}$, therefore $H(m^2_l/m^2_h)\approx m_l/m_h$ applies for the three
pairs of neutrino masses, yielding
\begin{equation}
b_1 \approx - \frac{1}{(4\pi)^2} 
\left( - m_{\nu,2} m_{\nu,3} f_1 + m_{\nu,1} m_{\nu,3} f_2 - m_{\nu,1} m_{\nu,2} f_3
 \right)
.
\end{equation}

c) \textbf{Inverted hierarchy.} In this case $m_{\nu,3} \ll m_{\nu,1} \approx 
m_{\nu,2}$. Now $H(m_{\nu,3}^2/m_{\nu,2}^2) \approx m_{\nu,3}/m_{\nu,2}$ and
$H(m_{\nu,1}^2/m_{\nu,2}^2) \approx 1/3 $, hence
\begin{equation}
b_1 \approx - \frac{1}{(4\pi)^2} 
\left( m_{\nu,2} m_{\nu,3} (f_1 - f_2) - \frac{1}{3} \Delta m_{21}^2 f_3
 \right)
.
\end{equation}

\subsection{\textsf{Leptons vs quarks and numerical estimates}
\label{subsec:7.4}
}

It is very instructive to compare the behavior found for leptons with that of
quarks. The formulas for quarks are those of type $I_a$. Similarly to
(\ref{eq:7.12}), for quarks 
\cite{GarciaRecio:2009zp}
\begin{equation}
g_{\rm CP}  = J_q \Delta_u \Delta_d I_{u d}
,
\end{equation}
however, the integral $I_{u d}$ (the quark analog of $I_{\nu e}$) is not so
smooth as for leptons. For quarks the relevant splitting between light and
heavy is rather $m_u, m_d, m_s \ll m_c, m_b, m_t$ which is superficially
similar to $m_{\nu,i} \ll m_e, m_\mu, m_\tau$, but while for the leptons the
light particles are all of type ``up'' (weak isospin $+1/2$), in the case of
quarks the light particles are of mixed up and down type.  This is relevant
for the coupling due to the structure of the Jarlskog invariant which controls
CP violation in the Kobayashi-Maskawa mechanism.  After separation of the
heavy particles (making use of \neq{7.15})
\begin{equation}
\Delta_u \Delta_d I_{u d}
\approx
\frac{1}{m_c^2} (m_s^2-m_d^2) I_q
,
\qquad
I_q \equiv \int \frac{d^4p}{(2\pi)^4} p^6 N_u^2 N_d^2 N_s^2
.
\end{equation}
As it turns out, as a consequence of IR divergences, the dimensionless
quantity $(m_s^2-m_d^2) I_q$ is not a continuous function of the masses at
$m_u=m_d=m_s=0$. The directional limit there is finite but it depends on how
it is taken (i.e., on the ratios $m_u/m_s$ and $m_d/m_s$). The natural choice
$m_u,m_d\to 0$ followed by $m_s \to 0$ gives $1/(4\pi)^2$ which is close to
the exact result.

For the Dirac type coupling of leptons,
\begin{equation}
\Delta_\nu \Delta_e I_{\nu e}
\approx
\frac{1}{m_e^4 m_\mu^2}
(m_{\nu,1}^2-m_{\nu,2}^2) (m_{\nu,2}^2-m_{\nu,3}^2) (m_{\nu,3}^2-m_{\nu1}^2)
I_\nu
.
\end{equation}
The factor $(m_{\nu,3}^2-m_{\nu,1}^2) I_\nu$ is once again finite but not
continuous in the limit $m_{\nu,i}\to 0$, nevertheless, due to the additional
factors $(m_{\nu,1}^2-m_{\nu,2}^2) (m_{\nu,2}^2-m_{\nu,3}^2)$, the full
expression has a well-defined (zero) limit as $m_{\nu,i}\to 0$.

The different behavior of quarks and leptons does not stems from differences
in the formulas, but rather from the different pattern of separation between
light and heavy particles in each case. In the case of quarks, the IR
divergences (a kind of chiral enhancement \cite{Smit:2004kh,
  Hernandez:2008db,GarciaRecio:2009zp,Salcedo:2011hy}) allows to have a
sizable CP violating effective action even for relatively small quark
masses. This idea was forwarded in \cite{Smit:2004kh} and first confirmed
explicitly in \cite{GarciaRecio:2009zp}.

The previous analysis does not directly extend to the couplings of Majorana
type, but we can see that also in this case the amount of CP violation in the
effective action induced by leptons is small as it depends on the small
neutrino masses. At least for the lowest dimensional operators, which are
those in $\Gamma_{4+2}$.  Nevertheless, this conclusion could change for the
couplings of higher dimensional operators. As the dimension of the operator
increases, the integrals become more UV convergent, but also more IR divergent
in the massless limit, and so more sensitive to the IR regime.

\begin{table}[h]
\centering
\begin{tabular}{|l|ccc|c|}
\hline
 & QD & NH & IH & Quarks
\\
\hline
$|a_1|$ & $ 7.0 \times 10^{-33} $ & $ 1.4 \times 10^{-31} $ &  $ 4.8 \times 10^{-32} $& $3.8 \times 10^{-4}$
\\
$|b_1|$ & $ 7.4 \times 10^{-11} $ & $ 2.0 \times 10^{-11} $ &  $ 1.2 \times
10^{-12} $ & ($1.2 \times 10^{-7} $)
\\
\hline
\end{tabular}\\
\caption{Upper bounds (maximal $|K^k_\gamma|$ are assumed) for the
  coefficients of Dirac type ($a_1$) and Majorana type ($b_1$) in the three
  neutrino scenarios: quasi degenerate (QD), normal hierarchy (NH), and
  inverted hierarchy (IH). The value $\bar{m}_\nu=0.1\,{\rm eV}$ has been
  adopted in $a_1$ for QD neutrinos. For quarks $|g_{\rm CP}|$ is shown
  assuming a maximal value of $J_q$. The same coupling using the experimental
  value of $J_q$ is shown below, between parenthesis. Units in ${\rm
    GeV}^{-2}$.}
\label{tab:9}
\end{table}
Since the precise values of the neutrino masses and the PMNS matrix, including
phases, are not known yet, the couplings $a_1$ and $b_1$ cannot be given
definite values. In order to have a feeling of the strength of CP violation
induced by leptons, we present in Table \ref{tab:9} estimates of $a_1$ (Dirac
type) and $b_1$ (Majorana type). These estimates are really upper bounds
obtained by taking $|K^k_\gamma|=1/4$ and $J_{\rm CP}=1/(6\sqrt{3})$, as well
as $\bar{m}_\nu= 0.1\,{\rm eV}$. For comparison the value of $g_{\rm CP}$ for
quarks is also given, with $J_q=1/(6\sqrt{3})$. The number in parenthesis is
$g_{\rm CP}$ using the experimental value $J_q = 3\times 10^{-5}$
\cite{Beringer:1900zz}.  In the Majorana case, all $|f_k|$ are estimated (or
rather bounded) as $\bar{f} \approx 1/(2m_e^4)$ and $|f_i-f_j| \approx 2
\bar{f}$.

The results shown in Table \ref{tab:9} indicate that couplings of the type
$I_a$ are much smaller than those of quarks.  It follows that the leptonic
corrections to the couplings of operators which already have a quark
contribution is extremely small. Note though that even these small values are
many orders of magnitude larger that the perturbative estimates discussed in
the Introduction, when these are applied to leptons (these give ratios as
small as $10^{-93}$.)

The typical values of couplings of Majorana type are about 20 orders of
magnitude larger than those of Dirac type. This indicates that CP violation
could serve as a test to discriminate the two types of neutrino masses,
provided the CP violating phases in the PMNS matrix are not too small.  The
Majorana couplings are smaller that those of quarks but these new couplings
are the only ones present for parity odd operators.

Before finishing this section we want to briefly discuss the structure of
the other couplings, $a_2$, $b_2$, $b_3$, $b_4$ and $b_5$ defined in
\neq{6.22}.

The relation similar to \neq{7.12} for $a_2$ is
\begin{equation}
a_2 = I^8_{a,1,1,2,3} =
J_{\rm CP} \Delta_\nu \Delta_e I^\prime_{\nu e}
,
\end{equation}
with
\begin{equation}
I^\prime_{\nu e} = 
\int \frac{d^4p}{(2\pi)^4} p^8
 \sum_{j=1}^3 N_{\nu,j}
 \prod_{i=1}^3 N_{\nu,i}^2 
\prod_{\alpha=1}^3 N_{e,\alpha}^2
.
\end{equation}
This quantity can be analyzed along the same lines as $a_1$ and similar
formulas are obtained (e.g., the factor $1/5$ in \neq{7.20} becomes $1/2$ for
$a_2$). So this coefficient will be of similar size as $a_1$ itself.

Rather than doing a detailed analysis of the coefficients it is instructive to
look for potentially large values by studying the possible IR divergences in
the integrals.

Consider first the integral $a_1=I^6_{a,1,1,2,2}$ for leptons in
the limit of small neutrino masses. The integral has many contributions of the
type 
\begin{equation}
I^6_{a,1,1,2,2} \sim \int d^4p \, p^6 N_{e,\alpha} N_{e,\beta}^2 N_{\nu,i}
N_{\nu,j}^2
\qquad \text{(leptons)}
\end{equation}
and by choosing different labels of type $\nu$ and $e$, we would want to
select the worst cases, i.e. the most IR divergent, ones. However, one can see
that this integral is IR finite as the neutrino masses go to zero.
The same conclusion follows for $a_2=I^8_{a,1,1,2,3}$.

If instead of leptons, the analysis of $I^6_{a,1,1,2,2}$ is carried out for
quarks, with the masses of $u$, $d$ and $s$ going to zero, the worse case is
\begin{equation}
I^6_{a,1,1,2,2} \sim \int d^4p \, p^6 N_c N_u^2 N_s N_d^2
\qquad \text{(quarks)}
\end{equation}
which is IR divergent at a logarithmic rate. Note that due to antisymmetry of
$I_a$ with respect to up-like and down-like (or neutrinos and charged leptons)
separately (see \neq{4.10a}) the two up-like ($c$ and $u$ above) and the two
down-like ($s$ and $d$ above) must be different. The presence of the IR
divergence causes the quarks to give a larger contributions than neutrinos in
type $I_a$ integrals.

Let us consider now the integrals of type $I_b$. In this case, the two
neutrino labels should still be different but the charged lepton labels can be
equal. For $b_1=I^2_{b,1,1,1,2}$ a typical contribution is
\begin{equation}
I^2_{b,1,1,1,2} \sim  m_{\nu,i} m_{\nu,j} \int d^4p \, p^2 N_e N_e N_{\nu,i} N_{\nu,j}^2
.
\end{equation}
The integral would be logarithmically IR divergent as the two neutrino masses
go to zero but this is multiplied by the product of the same masses yielding
an $O(m_\nu^2)$ final result. The analysis is similar for $b_2=
I^4_{b,1,1,1,3}$. For $b_3= I^4_{b,1,1,2,2}$, $b_4= I^6_{b,1,1,3,2}$ and $b_5=
I^6_{b,2,1,2,2}$ the integrals themselves are IR finite since there are
additional powers of $p^2$ but no additional neutrino propagators, as compared
to $b_1$. Therefore, the other $b_i$ couplings are not expected to be larger
than $b_1$. The typical scale of the couplings of type $b$ is set by the
factor $|\Delta m^2_{31}|/m_e^4 \approx 10^{-8} \, {\rm GeV}^{-2}$, while
$|\Delta m^2_{31}|^2 /(m_e^4 m_\mu^2) \approx 10^{-27} \, {\rm GeV}^{-2}$ sets
the scale in the couplings of type $a$. Compared to this, the scale of CP
violation for quarks is set by the factor $1/m_c^2 \approx 1 \, {\rm
  GeV}^{-2}$, in dimension six operators.

\section{\textsf{Summary and conclusions}}
\label{sec:8}

In this work we have undertaken a direct calculation of the couplings to CP
violating operators as induced by the leptonic loop in the Standard Model,
extended to include neutrino masses either of pure Dirac type or pure Majorana
type. The study is restricted to leading dimension operators, which have been
shown to be of dimension six, although four dimensional CP odd operators would
be obtained in the mixed Dirac-Majorana case (see \neq{4.11}). Our calculation
is exhaustive at dimension six. The results are collected in Tables
\ref{tab:2}-\ref{tab:8}. We have confirmed that CP would be violated in a two
generation scenario with Majorana neutrinos present.  Also, we show that CP
odd and P odd operators are produced at dimension six with Majorana neutrinos,
while only P even operators are produced for Dirac neutrinos.

An interesting result of the present study is the derivation of a Klein-Gordon
like operator, the operator $K$ (see \neq{4.23}), which describes the
propagation of leptons in the sectors $\langle \psi_R \bar{\psi}_L\rangle$ and
$\langle \psi_L \bar{\psi}_R \rangle$, also in the Majorana case, and which
produces the correct effective action in all sectors, CP odd and CP even, and
beyond the derivative expansion.

Two different mechanisms are present for Majorana neutrinos, one (type $I_a$)
is common to the Dirac neutrino case, and a new one (type $I_b$) is exclusive
of the Majorana case. The two types contribute simultaneously to the same set
of CP odd operators. We have shown that type $I_a$ contributions are identical
to those of the Dirac case when no $Z$ or Higgs are involved, but new terms
appear when these fields are present, and in particular, P can be violated in
the CP odd sector at dimension six. At variance with this, type $I_b$ terms
allow to break P at dimension six without requiring Higgs or $Z$ fields.

In the discussion of the new invariants of the PMNS matrix introduced by the
presence of Majorana masses, we have been able to identify two quadratic
relations among the nine invariants and a relation with the Jarlskog invariant
(see \neq{7.9}). Because only six parameters of the PMNS are effective in the
$I_b$ terms (the three angles, the Dirac phase and the two Majorana phases) it
follows that one more non-linear relation is missing among the nine invariants
but we have not been able to identify it.

The general expressions of the couplings, which come as momentum integrals
have been approximated in two typical cases, by exploiting the small neutrino
mass, yielding manageable formulas. Numerically we find that the couplings for
Dirac neutrinos, or more generally type $I_a$ terms, are many (about fifty or
sixty) orders of magnitude larger that perturbative estimates, but still much
smaller (by about a factor $10^{-28}$) than similar contributions from the
quark loop.  The reason for this difference is not so much the small mass of
the neutrinos but the fact that the pattern of breaking between light and
heavy fermions is different in quarks and leptons, as regard to weak isospin:
for quarks, the three lightest particles, $u$, $d$ and $s$, are of up and down
mixed type, whereas for leptons the three light neutrinos are of type up, and
this is relevant for the Jarlskog determinant and the IR structure of the
couplings. For the dimension six operators we have considered, the scale of CP
violation in couplings of type $I_a$ (Kobayashi-Maskawa mechanism) is
controlled by the factor $|\Delta m^2_{31}|^2 /(m_e^4 m_\mu^2) \approx
10^{-27} \, {\rm GeV}^{-2}$ for leptons and by $1/m_c^2 \approx 1 \, {\rm
  GeV}^{-2}$ for quarks.

The scale of CP violation for couplings of type $I_b$ (virtual lepton-number
violation) specific of Majorana neutrinos is set by the factor $|\Delta
m^2_{31}|/m_e^4 \approx 10^{-8} \, {\rm GeV}^{-2}$. Numerically, these
couplings are smaller than quark-induced terms only by about a factor
$10^{-7}$, assuming generic phases in the PMNS matrix. More importantly, some
of the operators induced by integration of the Majorana neutrinos are
exclusive for this mechanism. In particular, parity violating operators appear
at dimension six for Majorana neutrinos which are not present for Dirac
neutrinos. In principle, this opens the door to distinguish between Dirac and
Majorana neutrinos in precision tests involving CP violation provided the
relevant selection rules, for the appropriate operators, can be enforced.

The main limitation of the calculations of the present type regarding
phenomenology is the use of a strict derivative expansion, which implies an
expansion around zero four-momentum for the external fields. The lifting of
this restriction is a direction in which the present work could be
extended. Other directions include studying the case of mixed Dirac and
Majorana neutrino masses, which would allow to consider new operators of lower
dimension, the study of higher dimensional operators which have a different IR
behavior, and the inclusion of finite temperature effects, relevant for
baryogenesis scenarios.

\newpage

\appendix

\section{\textsf{Covariant and ordinary symbols}}
\label{app:A}

In \Eq{6.8s} there appear the covariant symbols of several operators. These
symbols are multiplicative with respect to $x$ but contain derivatives with
respect to $p_\mu$. These derivatives will be indicated as $\partial^p_\alpha =
\partial /\partial p_\alpha$. To compute the symbols we apply the relations
\neq{6.12}, using for each field its proper gauge connection, namely, $-qA_e$
with $q= \pm1, 0, 0$, for $W^\pm$, $Z$ and $\varphi$, respectively.

In general the covariant symbols are infinite series ordered by the number of
derivatives. For $\Gamma_{4+2}$ we need to retain two derivatives, counting
$D_{e,\mu}$, $Z_\mu$ and $\varphi_\mu$ as first order. The correct number of
$W$'s is already selected in \Eq{6.8s}.

\textbf{Derivatives and gauge and Higgs fields:}
\begin{equation}
\begin{split}
\bar{W}^\pm_\mu &= W^\pm_\mu + i W^\pm_{\alpha\mu}\partial^p_\alpha 
- \frac{1}{2}
W^\pm_{\alpha\beta\mu}\partial^p_\alpha \partial^p_\beta
 + O(D^3)
,
\\
\bar{Z}_\mu &= Z_\mu + i Z_{\alpha\mu}\partial^p_\alpha 
 + O(D^3)
,
\\
\bar{\varphi}_\mu &= \varphi_\mu + i \varphi_{\alpha\mu}\partial^p_\alpha 
 + O(D^3)
,
\\
\bar{D}_{e,\mu} &= i p_\mu + \frac{i}{2} F^e_{\alpha\mu} \partial^p_\alpha 
 + O(D^3)
,\\
\bar{D}^*_{e,\mu} &= i p_\mu - \frac{i}{2} F^e_{\alpha\mu} \partial^p_\alpha 
 + O(D^3)
,\\
\bar{D}_{e\pm,\mu} &= \bar{D}_{e,\mu} 
\pm (\bar{Z}_\mu + \bar{\varphi}_\mu)
,
\\
\bar{D}_{\nu,\mu} &= i p_\mu 
,
\\
\bar{D}_{\nu\pm,\mu} &= \bar{D}_{\nu,\mu} \pm (\bar{Z}_\mu - \bar{\varphi}_\mu) 
,
\\
\bar{F}^e_{\mu\nu} &= F^e_{\mu\nu} + O(D^3)
.
\end{split}
\end{equation}

\textbf{Mass terms:}
\begin{equation}
\begin{split}
\bar{m}_e^n &= m_e^n \left(
1 + n i \varphi_\alpha \partial^p_\alpha 
- \frac{1}{2}( n \varphi_{\alpha\beta} +  n^2 \varphi_\alpha \varphi_\beta ) 
\partial^p_\alpha \partial^p_\beta  + O(D^3)
\right)
,
\\
\bar{m}_\nu^n &= m_\nu^n \left(
1 + 2n i \varphi_\alpha \partial^p_\alpha 
- \frac{1}{2}( 2n\varphi_{\alpha\beta} + 4n^2 \varphi_\alpha \varphi_\beta )
\partial^p_\alpha \partial^p_\beta + O(D^3)
\right)
,
\end{split}
\quad
n=1,2,\ldots
\end{equation}

\textbf{Propagators:} There are various propagators, $G_e$, $G_e^{(2)}$,
$G_e^*$, $G_\nu$, and $G_\nu^\prime$. All of them follow the scheme
\begin{equation}
\bar{G} = N - N H N + N H N H N + O(D^3)
\end{equation}
where $N$ is of order zero and $H$ is at least of order one in
derivatives. For the various $N$
\begin{equation}
\begin{split}
N_e &= N_e^{(2)} = U^\dagger (m_e^2 + p^2 )^{-1} U
, \\
N_e^* &= U^T (m_e^2 + p^2 )^{-1} U^*
, \\
N_\nu &= N_\nu^\prime = (m_\nu^2 + p^2 )^{-1}
,
\end{split}
\end{equation}
while for the $H$:
\begin{equation}
\begin{split}
H_e &= \bar{m}_e^2 - \bar{\sD}_{e-} \bar{\sD}_e -m_e^2 - p^2
,
\\
H_e^{(2)} &= \bar{m}_e^2 - (\bar{\sD}_{e-} + 2 \bar{\thru\varphi} )
( \bar{\sD}_e  + 2 \bar{\thru\varphi}) - m_e^2 - p^2
, \\
H_e^* &= \bar{m}_e^2 - \bar{\sD}^*_e\bar{\sD}^*_{e+} - m_e^2 - p^2
, \\
H_\nu &= \bar{m}_\nu^2 - \bar{\sD}_{\nu+}\bar{\sD}_{\nu-} - m_\nu^2 - p^2
, \\
H^\prime_\nu &= \bar{m}_\nu^2 - (\bar{\sD}_{\nu-} - 2 \bar{\thru\varphi})
(\bar{\sD}_{\nu+} + 2 \bar{\thru\varphi}) - m_\nu^2 - p^2
.
\end{split}
\end{equation}
In the formulas with expressions valid to all orders one has to expand in
derivatives dropping terms of $O(D^3)$. E.g.,
\begin{equation}
\begin{split}
H^\prime_\nu &= 
m_\nu^2 \left(
4 i \varphi_\alpha \partial^p_\alpha 
- (  2 \varphi_{\alpha\beta} +  8 \varphi_\alpha \varphi_\beta ) 
\partial^p_\alpha \partial^p_\beta 
\right)
\\ &
- i [\thru{p}, \thru{Z} + \thru{\varphi} ]
+ [ \thru{p}  , \gamma_\mu ( Z_{\alpha\mu} + \varphi_{\alpha\mu} ) \partial^p_\alpha ]
+ ( \thru{Z} + \thru{\varphi} )^2
 + O(D^3)
.
\end{split}
\end{equation}

Alternatively one can use the method of {\em ordinary symbols}
\cite{Nepomechie:1984wt,Salcedo:1994qy}. In this case, the equation similar to
\neq{6.12a} is
\begin{equation}
\Tr \, f(D,M) = \int \frac{d^dx d^dp}{(2\pi)^d} \,\tr f(D + ip, M)
.
\label{eq:6.12b}
\end{equation}
After integration over $p_\mu$ all covariant derivatives appear only in the
form $[D_\mu,~]$, but unlike the case of covariant symbols, these explictly
covariant combinations have to be obtained by hand (essentially moving the
$D$'s to the right using $D_\mu X=X D_\mu + \hat{D}_\mu X$). This makes this
method less systematic. It should be noted that the cyclic property of the
trace can be freely applied to writing the starting pseudodifferential
operator, $f(D,M)$, without changing the UV convergent contributions in the
final result, however, in general such freedom is not justified for the
different terms in $\tr f(D + ip, M)$.

\acknowledgments
This work has been supported by Spanish DGI (FIS2011-24149) and Junta de
Andaluc\'{\i}a grant FQM-225.


\providecommand{\href}[2]{#2}\begingroup\raggedright\endgroup

\end{document}